\documentclass[%
reprint,
 amsmath,amssymb,
aps,
onecolumn,
]{revtex4-2}

\usepackage{graphicx}
\usepackage{dcolumn}
\usepackage{bm}

\usepackage{siunitx}
\usepackage{tabularx}
\usepackage{tikz}

\newcommand{\mt}[1]{\mathrm{#1}}
\newcommand{\bs}[1]{\boldsymbol{#1}}

\newcommand\Tstrut{\rule{0pt}{2.6ex}}         
\newcommand\Bstrut{\rule[-0.9ex]{0pt}{0pt}}   

\newcommand{\ft}[1]{\widetilde{#1}}
\newcommand{\fh}[1]{\widehat{#1}}

\usepackage{xcolor}
\definecolor{purple}{RGB}{191,0,191}
\definecolor{orange}{RGB}{255,127,14}
\definecolor{green}{RGB}{44,160,44}
\definecolor{blue}{RGB}{31,119,180}
\definecolor{purple2}{RGB}{148, 103, 189}
\definecolor{brown}{RGB}{140, 86, 75}
\definecolor{magenta}{RGB}{227, 119, 194}

\definecolor{black}{RGB}{0,0,0}
\definecolor{white}{RGB}{255,255,255}

\begin{document}

\preprint{APS/123-QED}

\title{Study of the energy convergence of the Karhunen--Loève decomposition applied to the large-eddy simulation of a high-Reynolds-number pressure-driven boundary layer}

\author{Pieter Bauweraerts}
 \email{pieter.bauweraerts@kuleuven.be}
\author{Johan Meyers}%
 \email{johan.meyers@kuleuven.be}
\affiliation{%
KU Leuven, Mechanical Engineering,
Celestijnenlaan 300A, B3001 Leuven, Belgium
}%

\date{\today}

\begin{abstract}
We study the energy convergence of the Karhunen--Loève decomposition of the turbulent velocity field in a high-Reynolds-number pressure-driven boundary layer as function of the number of modes. An energy-optimal Karhunen--Loève (KL) decomposition is obtained from wall-modelled large-eddy simulations at `infinite' Reynolds number. By explicitly using Fourier modes for the horizontal homogeneous directions, we are able to construct a basis of full rank, and we demonstrate that our results have reached statistical convergence. The KL-dimension, corresponding to the number of modes per unit volume required to capture $90\%$ of the total turbulent kinetic energy, is found to be $\SI{2.4e5}{|\Omega|/H^{3}}$ (with $|\Omega|$ the domain volume and $H$ the boundary layer height). This is significantly higher than current estimates, which are mostly based on the method of snapshots. In our analysis, we carefully correct for the effect of subgrid scales on these estimates. 
\end{abstract}

\maketitle


\section{\label{sec:Introduction} Introduction}
Low dimensional models describing the dynamics of the atmospheric boundary layer (ABL) have important applications, e.g., ranging from dispersion of pollutants, to predicting the power output of a wind turbine, to controlling the turbulence in wind farms for enhanced power production. An assessment of the required dimensionality of a reduced order system can be made by studying the system dynamics, typically done via the Kaplan--Yorke (KY) definition of the dimension of the attractor, which is based on the Lyapunov exponents of the dynamical system \cite{kaplan1979chaotic}. This becomes increasingly difficult for large scale systems. Alternatively the Karhunen--Loève (KL) decomposition (also known as proper orthogonal decomposition (POD), empirical orthogonal functions (EOF) or principal component analysis (PCA)) can be used, which generates modes which are well known to be energy optimal, in the sense that of all possible mode sets they capture on average the most energy \cite{pearson1901liii}. The KL-modes are often used to identify structures in turbulent flows, and employed as a basis for a reduced order system. An assessment of the number of modes required is the so called KL dimension, introduced into the fluid dynamics community by \textcite{sirovich1989chaotic}. This is defined as: ``\textit{the number of actual
eigenfunctions required so that the captured energy is at least $90\%$ of the total (as measured by the energy norm) and that no neglected mode, on average, contains more than $1\%$ of the energy contained in the principle eigenfunction mode.}'' This dimension has previously been found to be of the same order of magnitude as the KY-dimension \cite{sirovich1991computational}. The dimensionality of turbulent flows is typically significantly lower than the amount of degrees of freedom (DOF) $\sim\int_{\Omega}\eta^{-3}\, \mt{d}\bs{x}$, with $\eta$ the local Kolmogorov scale and $\Omega$ the domain, due to the prevalence of larger scale structures containing the majority of the energy. Moreover, at asymptotically high Reynolds numbers, such as encountered in the ABL, we expect a finite KL dimension that becomes independent of the Reynolds number. In the current work, we use large-eddy simulations (LES) of a high-Reynolds number pressure driven boundary layer (PDBL) to study the energy convergence of this type of flows. We show that the KL dimension is up to three orders of magnitude higher than what is commonly reported in literature. 

The KL-modes $\phi_i$ ($i=1\cdots 3$) are found as the eigenfunctions of the two-point covariance tensor \cite{berkooz1993proper}
\begin{align} \label{eq:Fredholm}
\langle R_{ij}(\bs{x},\bs{\breve{x}})\phi_j(\bs{\breve{x}})\rangle_{\breve{\bs{x}}} = \lambda\phi_i(\bs{x}),
\end{align}
with $R_{ij}(\bs{x},\bs{\breve{x}}) = \langle u_i'(\bs{x},t)u_j'(\breve{\bs{x}},t)\rangle_t$ the two-point covariance tensor, $u_i'(\bs{x},t)=u_i-\langle u_i \rangle_t$ are the velocity fluctuations around the local mean and $\langle \cdot \rangle_{t}$ the averaging operator over time $t$, and used similarly for other variables. Further, $\lambda$ are the eigenvalues, which represent twice the space-time averaged turbulent kinetic energy (TKE) captured by the eigenfunction, i.e. $\lambda = \langle\langle u_i'\phi_i\rangle_{\bs{x}}^2 \rangle_{t}/\langle\phi_j\phi_j \rangle_{\bs{x}}$. The eigenvalues are numbered and ordered in decreasing order $\lambda_{k}\geq\lambda_{k+1}$. The averaged TKE captured by the largest $n$ eigenvalues $K_n$ is given by \cite{berkooz1993proper}
\begin{equation}
K_{n} = \frac{1}{2}\sum_{k=1}^{n}\lambda_k.
\end{equation}
For convenience, we introduce the unresolved energy fraction 
\begin{equation}\label{eq:energy_frac_def}
\mathcal{E}_{n} =  1-\frac{K_{n}}{K},
\end{equation}
where $K=K_{\infty}=\frac{1}{2}\langle u_i' u_i'\rangle_{x,t}$ is the total average TKE.
Based on this, the KL-dimension $D_{\mt{KL}}$ is expressed mathematically as
\begin{align} \label{eq:KL_def}
D_{\mt{KL}} = \underset{n}{\text{arg min}}\ n\text{, subject to }\mathcal{E}_n \geq 0.1\text{ and } \lambda_{n+1}\leq 0.01\lambda_1.
\end{align}
Note that the final constraint is typically not active, due to the high-dimensional nature of turbulent flows. The KL-dimension is case dependent, and is a function of the dimensionless groups obtained from main flow parameters, such as friction velocity, BL-height, viscosity, surface roughness, and the domain shape. 

After discretization the two-point covariance tensor $\bs{R}$ has $3N\times3N$ elements, where $N$ is the amount of grid points. For a typical high-Reynolds-number turbulent boundary-layer simulation, $N$ in the order of $10^7-10^{10}$ points, such that directly solving the eigenvalue problem (\ref{eq:Fredholm}) is too big to handle with current computational resources. A popular strategy for transforming this into a tractable problem is the so-called method of snapshots \cite{sirovich1987turbulence}, where the the spatial eigenvalue problem is transformed into an eigenvalue problem in time. This is obtained by multiplying (\ref{eq:Fredholm}) by $u_i'(\bs{x},\breve{t})$ and averaging over $\bs{x}$, and leads to \cite{sirovich1987turbulence}
\begin{align} \label{eq:Fredholm_T}
\langle \rho(\breve{t},t)\, \psi(t) \rangle_{t} = \lambda\psi(\breve{t}),
\end{align}
where $\rho(t,\breve{t})= \langle u_i'(\bs{x},t)u_i'(\bs{x},\breve{t})\rangle_{\bs{x}}$ is the spatially averaged time covariance between two fields, and $\psi(t)=\langle \phi_i(\bs{x})u_i'(\bs{x},t)\rangle_{\bs{x}}$ is the projection of the velocity field fluctuations on the POD mode $\phi_i(\bs{x})$. A similar relationship can be found for transforming the time modes $\psi(t)$ to the spatial modes
$\phi_i(\bs{x}) = \langle \psi(t) u_i'(\bs{x},t) \rangle_{t}/\lambda$.
After discretization (\ref{eq:Fredholm_T}) becomes an eigenvalue problem of size $N_{\mt{s}}$, with $N_{\mt{s}}$ the amount of samples, typically of the order $10^3$-- $10^4$. This strategy is applicable to general flow geometries, but the rank is thereby capped by the number of samples, and the eigenvalue spectrum has been shown to converge very slowly for large scale problems \cite{duggleby2008structure}.

An alternative approach is available if the problem exhibits homogeneous directions, which is often the case for canonical flow cases studied in turbulence. In case of a PDBL, both horizontal directions are homogeneous. In this case, the two-point covariance tensor can be rewritten as 
 \begin{equation}
 R_{ij}(\bs{x},\breve{\bs{x}}) \rightarrow  R_{ij}(x_1-\breve{x}_1,x_2-\breve{x}_2, x_3,\breve{x}_3).
 \end{equation}
It is easily shown that the POD-modes correspond to Fourier modes in these directions, i.e. $\phi_i(\bs{x})=\exp(\mt{i}(k_1x_1+k_2x_2))\fh{\phi}_{i}^{(\bs{k})}(\breve{x}_3)$, where $\bs{k}=[k_1,k_2]$ is the horizontal wave vector, and $\fh{\phi}_{i}^{(\bs{k})}(x_3)$ the horizontally Fourier transformed POD mode. The large scale eigenvalue problem can then be replaced by a smaller scale eigenvalue problem per wave number (see e.g. \cite{berkooz1993proper})
 \begin{align}
	\langle \fh{R}_{ij}(\bs{k},x_3,\breve{x}_3) \fh{\phi}_{i}^{(\bs{k})}(\breve{x}_3) \rangle_{\breve{x}_3} = \lambda_{\bs{k}}\fh{\phi}_{j}^{(\bs{k})}(x_3), 
	\end{align}
with $\fh{R}_{ij}(\bs{k},x_3,\breve{x}_3) = \langle \fh{\ft{u}}\vphantom{u}'^{*}_i(\bs{k}, x_3) \fh{\ft{u}}\vphantom{u}'_j(\bs{k}, \breve{x}_3)\rangle_t$. After discretization leads to a problem of size $3N_3$ per wave number, where $N_3$ is the amount of grid cells in the vertical direction. This results in a set of eigenvalues $\lambda_{\bs{k},m}$, with $m=1,\ldots, 3N_3$ per wave number. The contributions of the different wave numbers can then be brought together again to obtain a complete basis, and ordering the eigenvalues regardless of the originating wave number from largest to smallest.
	
The KL-dimension has already been determined for canonical flow cases at low Reynolds/Rayleigh numbers. An overview of studies is given in Table~\ref{tab:KLdimension}, including reported KL dimensions, etc.  The KL-dimension for turbulent channel flow has been determined in Ref. \cite{ball1991dynamical}, while in Ref. \cite{webber1997karhunen} a minimum flow unit was studied and compared to a larger test case, identifying a linear increase in KL-dimension with dimension, in correspondence with the KL-dimension being an extensive property. Refs. \cite{sirovich1991propagating,iwamoto2002reynolds} on the other hand found a strong increase in dimensionality with the Reynolds number, while Ref. \cite{housiadas2005viscoelastic} considered the influence of visco-elasticity. Other flow cases considered are Couette flow \cite{smith2005low}, Rayleigh--Bénard convection \cite{ciliberto1991estimating, sirovich1991computational}, turbulent pipe flow \cite{duggleby2007dynamical}, and a turbulent boundary layer \cite{cardillo2013dns}.

POD studies considering the high-Reynolds atmospheric boundary layer, simulated using LES, are numerous. A non-exhaustive overview of case set-ups and reported KL dimensions is also provided in Table~\ref{tab:KLdimension}. Three-dimensional POD of the ABL using snapshot POD approach are performed in Refs. \cite{verhulst2014large, ali2017turbulence, zhang2019characterizing}. A comparison of  POD in 2D planes, significantly reducing the dimensions, but losing part of the 3D structures, is performed in Refs. \cite{shah2014very,andersen2014reduced,andersen2017turbulence, bastine2015towards,bastankhah2017wind, bastine2018stochastic}. In Ref.~\cite{huang2009analysis} both 1D and 3D POD computations were performed, where homogeneity assumptions are used for the latter. For the 1D case, the convergence of LES and direct numerical simulation (DNS) is compared. A slower convergence of the eigenvalues was found for LES, which was attributed to the Reynolds number (as was demonstrated in Ref. \cite{iwamoto2002reynolds}).  Other studies only considered a few dominant modes \cite{esau2003coriolis,keith1996empirical}. Note that studies \cite{verhulst2014large, ali2017turbulence, zhang2019characterizing} considered the impact of wind turbines in an atmospheric boundary layer, such that the Fourier approach is no longer applicable, and only discrete translational symmetry can be used in periodic directions to extend the snapshot base. Next to LES studies, also experimental wind tunnel studies considering 2D POD modes using the spectral approach \cite{finnigan2000wind}, and using the method of snapshots \cite{hamilton2015wind} are found. For reference, the simulation case considered in the current manuscript is also shown in Table~\ref{tab:KLdimension}, with a KL dimension that is significantly larger than other reported high-Reynolds-number boundary layers. In the remainder of the manuscript, this case and the determination of the resulting KL dimension is carefully documented.

In the current work, we consider LES of a rough-wall pressure-driven turbulent boundary as a substitute for a neutral atmospheric boundary layer. This approach has been often used for LES studies of the neutral ABL \cite{shaw1992large,porte2000scale, bou2004large,huang2009effects,chamecki2009large, lu2010modulated,calaf2010large}, and is known to represent statistics in the logarithmic layer very well. Simulations are based on wall-modelled LES, using a wall-stress model, and direct effects of viscosity are neglected, so that all dissipation is handled by the subgrid-scale model, approaching effectively the limit of an `infinite' Reynolds number. Analytical expressions for the modes in the horizontal homogeneous directions are used, such that a complete KL basis and the corresponding eigenvalues can be determined. We focus on the eigenvalue spectrum, and more precisely on the convergence. The structure of the associated dominant modes are already extensively reported in the ABL studies summarized above, and are not repeated in the current study. We further identify the KL dimension, i.e. the number of modes required to represent 90\% of the energy.

The paper continues by giving a brief overview of our case setup, LES model, discretization, and the sampling specifications. Subsequently, the results are presented, considering both the statistical convergence of the results and aspects of dimensionality. Finally, the conclusions are summarized.

\begin{table}
	\begin{tabular*}{\textwidth}{@{\extracolsep{\fill}}l l l l l l l l l}
		\hline
		Case & $Re_\tau$  &\Tstrut $D_{\mt{KL}}$ & $L_1/H$ & $L_2/H$  & $L_3/H$ &$d_{\mt{KL}}$ $\lbrack H^{-3} \rbrack$& Method & Reference \\
		\hline
		DNS-CH \Tstrut& $80$ & $380$ & $1.6\pi$ &$1.6\pi$& $2$& $15.40$ & $F-SC$ & \textcite{ball1991dynamical}\\
		DNS-CH \Tstrut  &$110$& $13452$\footnote{\label{fn:tab}The reported values did not take into account the degeneracy of the modes to calculate the dimensionality, therefore the reported dimension was multiplied by 4 as an approximation/upper boundary.} & $5\pi$&$2\pi$ & $2$& $68.16$ & $F-SC$ & \textcite{iwamoto2002reynolds} \\ 
		DNS-CH \Tstrut & $125$ &$4186$ & $5$& $5$& $2$& $83.72$ & $F-SC$ & \textcite{sirovich1991propagating}  \\
		DNS-CH \Tstrut  & $136$&$658$ & $\pi$& $0.3\pi$& $2$& $111.12$ & $F-SC$ & \textcite{webber1997karhunen} \\
		DNS-CH \Tstrut  & $180$& $18920^a$& $9$&$4.5$&$2$ & $233.6$ & $F-SC$ & \textcite{housiadas2005viscoelastic} \\
		DNS-CH \Tstrut  &$300$& $36520^a$ & $2.5\pi$&$\pi$& $2$& $740.04$ & $F-SC$ & \textcite{iwamoto2002reynolds}\\
		\hline
		\hline
		  LES-ABL\Tstrut & $\infty$& $-$& $30$& $30$& $1$& $-$ & $F-SC$& \textcite{keith1996empirical} \\
		 LES-ABL\Tstrut&$\infty$ & $-$& 8 & 8& 1& $-$ & $F-SC$ & \textcite{esau2003coriolis}\\
		 LES-ABL \Tstrut&$\infty$& $10^3$ & $2\pi$&  $\pi$& $1$& $50.66$  &$TC$ & \textcite{ali2017turbulence} \\
		\hline
		\hline
		  LES-PDBL\Tstrut&$\infty$& $-$& $2\pi$ & $2\pi$& 1& $-$ & $F-SC$& \textcite{huang2009analysis}\\
		LES-PDBL  \Tstrut&$\infty$& $4\times 10^3$& $\pi$ & $\pi$& $1$ & $405$ &  $TC$ & \textcite{verhulst2014large}\\
		 LES-PDBL \Tstrut&$\infty$& $3\times10^3$& $4\pi$ & $2\pi$ & $1$  & $38$ & $TC$ & \textcite{zhang2019characterizing}\\
		LES-PDBL \Tstrut&$\infty$ & $9.8\times10^6$& $42$ & $12$ & $1$ & $240000$ & $F-SC$ & Current manuscript\\
		\hline
		\hline
	\end{tabular*}
	\caption{Comparison of different 3D KL studies. $D_{\mt{KL}}$ is the total Karhunen-Loève dimension, $d_{\mt{KL}}$ is the same but normalized by the simulation volume (expressed in units of boundary layer height $H$). The abbreviations under `Case' are: CH~(channel flow), ABL~(atmospheric boundary layer, including Coriolis forces and/or stability effects), PDBL~(pressure driven boundary layer). The abbreviations under `Method' are: F~(Fourier), SC~(space correlation), and TC~(time correlation).}
	\label{tab:KLdimension}
\end{table}

\section{Case study and simulation methodology}
\begin{table}
	\begin{tabular}{l l l c}
		\hline 
		Domain size\Tstrut      &$L_1\times L_2  \times L_3$  &$42H\times 12H\times H$  \\
		Grid size & $N_1\times N_2 \times N_3$& $2800 \times 800 \times 200 $\\
		Cell size & $\Delta_1 \times\Delta_2 \times\Delta_3 $& $0.015H\times0.015H\times0.005H$ \\
		\\
		Roughness length\Tstrut      &$z_{0}/H$ & $\SI{2e-7}{}$\\
		\hline
	\end{tabular}
	\caption{Summary of the simulation grid setup and simulation parameters.}
	\label{tab:SimSetup}
\end{table}

\begin{figure}
	\centering
	\includegraphics[width=1.0\textwidth]{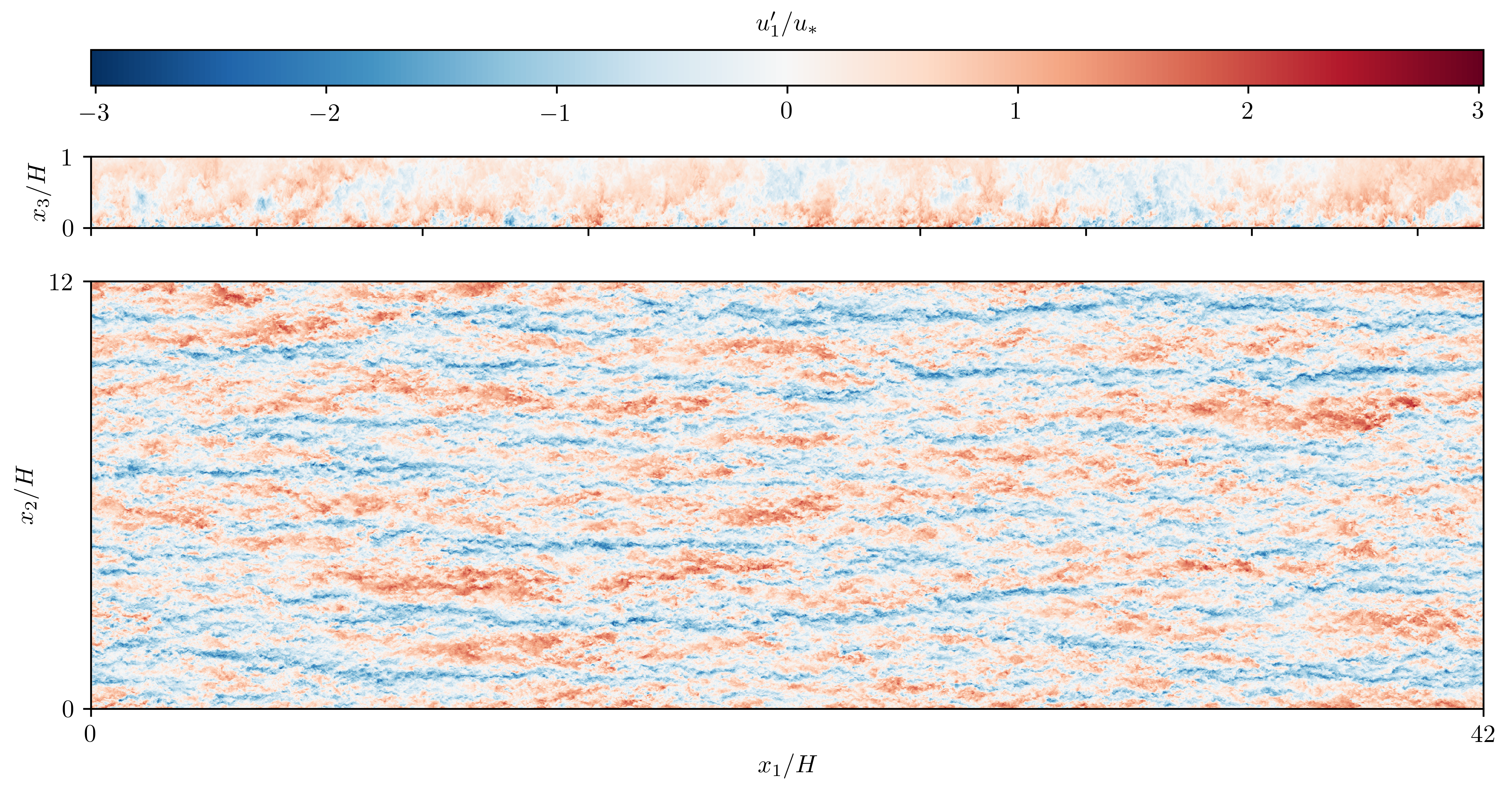}%
	\caption{ The top and bottom figure respectively represent a $x_1$--$x_3$ section at $x_2=L_2/2$ and a $x_1$--$x_2$ section at $x_3=0.1H$ of an instantaneous streamwise velocity field.}
	\label{fig:velocity_field}
\end{figure}

In this case study we consider a pressure driven boundary layer, a summary of the simulation parameters is given in Table \ref{tab:SimSetup}, and a snapshot of the flow field is given in Fig. \ref{fig:velocity_field}. We consider a relatively large domain of $42H\times12H\times H$ to avoid spurious influence of the periodic boundary conditions on the two-point velocity covariance tensor, which is known to extend up to $10H$ for the streamwise velocity component \cite{fang2015large, sillero2014two}. The surface roughness length $z_{0}$ is chosen such that for a BL height of $H=\SI{1000}{m}$, we get a value of $\SI{2e-4}{\metre}$, which is, e.g., typical for offshore conditions. 
The code used for this study has been extensively documented in past studies, see e.g. Ref. \cite{meyers2011error,munters2016shifted} for further details. In the horizontal directions we use periodic boundary conditions, in the vertical directions we use impermeability in combination with a wall-stress model \cite{bou2005scale} at the bottom wall and zero stress at the top.  As a subgrid-scale model, we use a classical Smagorinsky model \cite{smagorinsky1963general}, combined with wall-damping close to the wall \cite{Mason1992}. ABLs occur at very high Reynolds number, such that the effect of the kinematic viscosity on the resolved flow can be neglected. In this way our flow becomes Reynolds number independent and can be interpreted as the asymptotic behaviour at infinitely/very high  Reynolds number \cite{stevens2014large}. The horizontal directions use a Fourier spectral discretization, dealiased using the 3/2 rule (see e.g. \cite{canuto1988spectral}). For the vertical direction we employ a fourth-order energy conservative scheme  \cite{verstappen2003symmetry}. For the time integration we use an explicit fourth order Runge-Kutta method combined with a 0.4 Courant--Friedrichs--Lewy number limit on the time step. In order to speed up the simulations, the equations are solved in a frame of reference moving at approximately half the maximum flow speed $\sim9.5H/u_*$ in the streamwise direction, allowing for a doubling of the stability time step.

The flow field is sampled every $\SI{3.16e-3}{H/u_*}$ and the reflection symmetry of the equations in the spanwise directions is used to artificially double the sample size \cite{sirovich1987turbulenceII}. Note that for small wave numbers, subsequent samples remain correlated, such that the effective sample size will be smaller, and dependent on the considered wave numbers. A total of $8200$ samples is generated, leading to an averaging time of $12.97H/u_*$ time units.

\section{Results}
In order to carefully establish the KL dimension of our current simulation set-up, we first investigate in \S\ref{ss:convergence_KL_dimension} the effect on the KL spectra of the number of snapshots in computing the two-point covariance tensor, verifying sufficient convergence of the time average. Moreover, a first estimate of the KL dimension is established. Subsequently, in \S\ref{ss:spectrum_KLdim} the resulting KL spectra are further discussed, verifying their expected physical behaviour. Finally, in \S\ref{ss:CGcomp} we adapt the estimate of our KL dimension by taking into account the effect of the subgrid scales in our LES.

\subsection{Sampling time: convergence of the results} \label{ss:convergence_KL_dimension}
\begin{figure}
	\centering
	\includegraphics[width=1.0\textwidth]{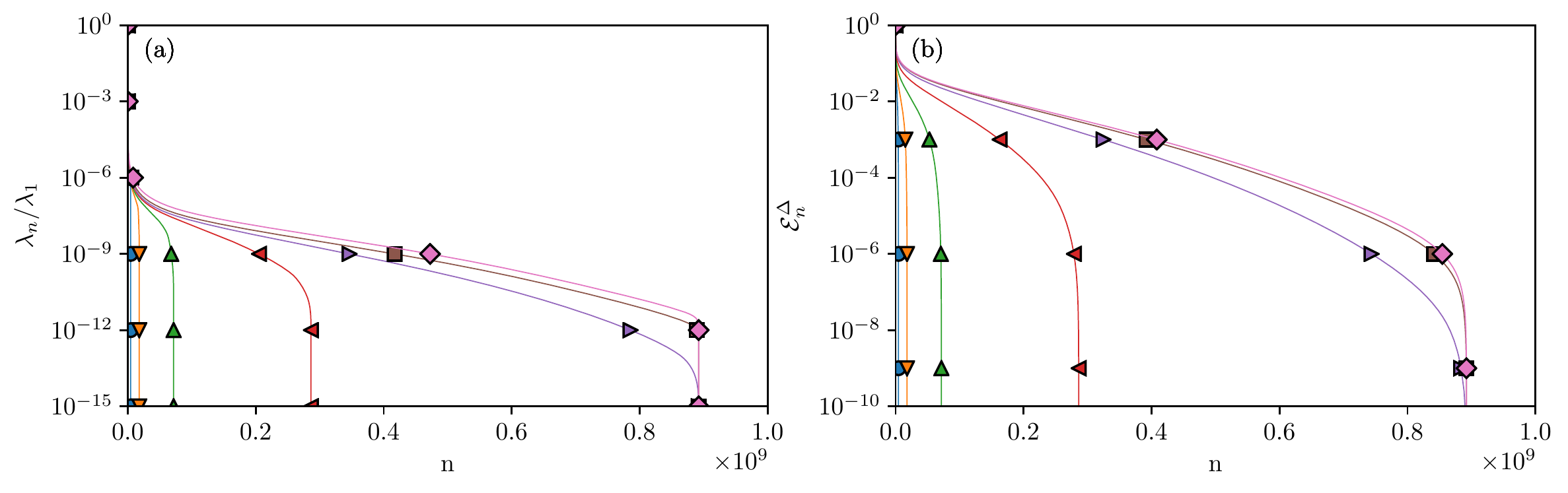}%
	\caption{ (a) convergence of the eigenvalues $\lambda_n$ (b) residual energy $\mathcal{E}_n^{\Delta}$ as a function of the index number $n$. The lines are computed using different amount of samples: (2,\protect\tikz{\protect\filldraw[draw=black,fill=blue] (0,0) circle [radius=0.1cm];}), (8,\protect\tikz{\protect\filldraw[draw=black,fill=orange] (0,0.2cm) --
			(0.2cm,0.2cm) -- (0.1cm,0.0cm) -- (0,0.2cm);}), (32,\protect\tikz{\protect\filldraw[draw=black,fill=green] (0,0.0cm) --
			(0.2cm,0.0cm) -- (0.1cm,0.2cm) -- (0,0.0cm);}), (128,\protect\tikz{\protect\filldraw[draw=black,fill=red] (0,0.1cm) --
			(0.2cm,0.2cm) -- (0.2cm,0.0cm) -- (0,0.1cm);}), (512,\protect\tikz{\protect\filldraw[draw=black,fill=purple2] (0.2cm,0.1cm) --
			(0.0cm,0.0cm) -- (0.0cm,0.2cm) -- (0.2cm,0.1cm);}), (2048,\protect\tikz{\protect\filldraw[draw=black,fill=brown] (0.0cm,0.0cm) --
			(0.2cm,0.0cm) -- (0.2cm,0.2cm) -- (0.0cm,0.2cm) -- (0.0cm,0.0cm);}) and (8192,\protect\tikz{\protect\filldraw[draw=black,fill=magenta] (0.1cm,0.0cm) --
			(0.2cm,0.1cm) -- (0.1cm,0.2cm) -- (0.0cm,0.1cm) -- (0.1cm,0.0cm);})}
	\label{fig:eigenvalue_convergence}
\end{figure}

Before discussing the convergence of the eigenvalues as a function of number of snapshots, we first introduce a further definition. Since we are performing LES, we do not formally know the total turbulent kinetic energy $K$, since a fraction of the kinetic energy is in the subgrid scales (see \S\ref{ss:CGcomp} for further discussion). Therefore, we introduce the resolved TKE, $K_{\Delta}=\frac{1}{2}\langle\ft{u}_i'\ft{u}_i'\rangle_{\bs{x},t}$, and further also
\begin{equation} \label{eq:energy_frac_fil}
	\mathcal{E}_n^{\Delta} = 1-\frac{K_{n}}{K_{\Delta}},
\end{equation}
which is the fraction of LES TKE resolved by the first $n$ POD modes. 

In Fig. \ref{fig:eigenvalue_convergence}, we show the KL spectra and $\mathcal{E}_n^{\Delta}$ for different numbers of samples in the calculation of the two-point covariance tensor. It is observed that starting from $2048$ samples upward, the shape of the different curves becomes almost independent of the amount of samples. The eigenvalue curves show that the amount of non-zero eigenvalues grows linearly with the amount of samples. For higher sample numbers an abrupt change occurs at around $2/3$ of the total amount of eigenvalues. This results from the rank of the spectral correlation matrix, which is limited by the amount of samples~$N_{\mt{s}}$. The POD basis will span the whole solenoidal space provided $N_{\mt{s}} > 2N_3$, in which case the last third of the eigenvalues are zero. This is a result from the fact that the two-point correlation tensor is based on solenoidal vector fields.

\begin{figure}
	\centering
	\includegraphics[width=0.4\textwidth]{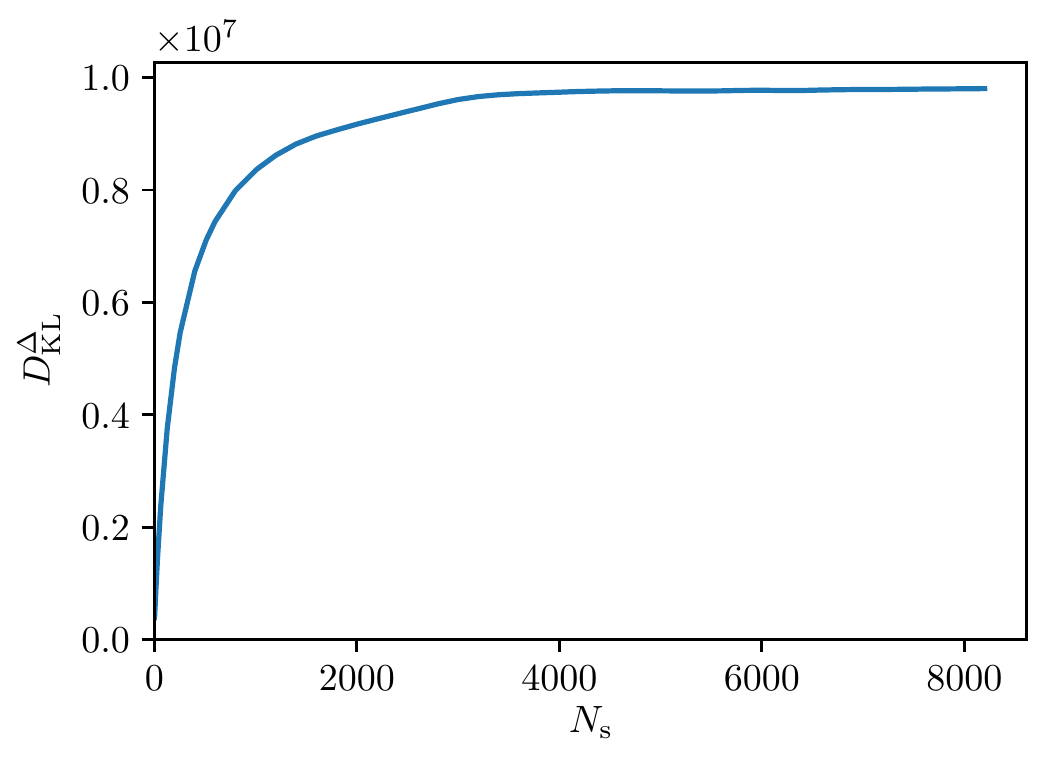}%
	\caption{KL-dimension $D_{\mt{KL}}^{\Delta}$ as a function of the amount of samples $N_{\mt{s}}$. }
	\label{fig:KL_dimension}
\end{figure}

The KL-dimension is the amount of POD methods necessary to capture $90\%$ of energy on average, and can be determined from the unresolved energy (see Fig. \ref{fig:eigenvalue_convergence} (b)). The influence of the amount of samples on the KL-dimension is shown in Fig. \ref{fig:KL_dimension}. After an  initial monotonous increase up to around $4000$ samples (corresponding to a total averaging time of $\SI{6.3}{H/u_*}$), $D_{\mt{KL}}^{\Delta}$ reaches a steady state value of $\SI{9.8e6}{}$. The superscript $\Delta$ is again added in the notation to indicate that results are based on a filtered velocity field $\ft{\bs{u}}$ in the LES, and do not account for possible subgrid energy (see \S\ref{ss:CGcomp} for more discussion). The KL-dimension is known to be an extensive property \cite{webber1997karhunen}, such that the dependence on the horizontal extent of the domain can be eliminated by dividing $D_{\mt{KL}}^{\Delta}$ by the non-dimensional volume $|\Omega'|=|\Omega|/H^3$, i.e. $d_{\mt{KL}}^{\Delta} =D_{\mt{KL}}^{\Delta}/|\Omega'|.$ Using the aforementioned value of $D_{\mt{KL}}^{\Delta}$, we obtain $d_{\mt{KL}}^{\Delta}=\SI{1.9e-4}{}$ or equivalently $D_{\mt{KL}}^{\Delta}=\SI{1.9e-4}{|\Omega'|}$. This is two to three orders of magnitude bigger than the numbers found by previous studies using snapshot POD of the ABL (see Table \ref{tab:KLdimension}). The difference can be explained by the slow convergence of the method of snapshots for high dimensional systems \cite{duggleby2008structure}. The substantial increase of KL-dimension with Reynolds number was already demonstrated in Ref. \cite{iwamoto2002reynolds}, for channel flows of $Re_\tau=180$ and $Re_\tau=300$, also see Table \ref{tab:KLdimension}. In Ref. \cite{huang2009analysis} a similar increase was found in the comparison of 1D vertical POD between DNS and LES. Finally, we note that, since the TKE is a large-scale property of turbulence, we expect at high Reynolds numbers, that the KL dimension becomes asymptotically independent of the Reynolds number, and similarly of  $H/\Delta$ in LES (see also \S\ref{ss:CGcomp} for a further discussion on the estimation of this asymptotic KL dimension).

\subsection{Eigenvalue spectrum} \label{ss:spectrum_KLdim}
\begin{figure}
	\centering
	\includegraphics[width=\textwidth]{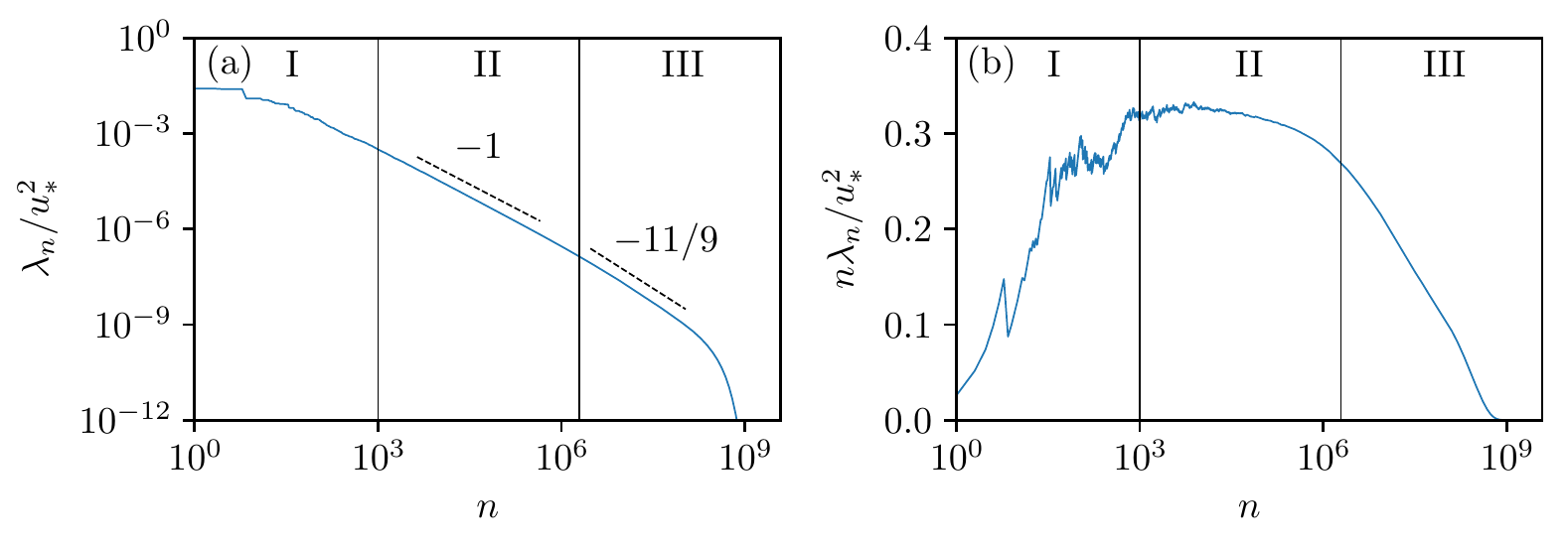}%
	\caption{(\protect\tikz[baseline=-0.8ex]\protect\draw [blue,thick] (0,0) -- (0.8,0);) (a) Eigenvalues $\lambda_n/u_*^2$ (b) premultiplied eigenvalues $n\lambda_n/u_*^2$ as a function of the index number $n$. (\protect\tikz[baseline=-0.8ex]\protect\draw [black,thick,dash pattern={on 2pt off 1pt}] (0,0) -- (0.8,0);) $n^{-1}$ and $n^{-11/9}$ scaling. The figure is suggestively subdivided in an inactive range (I), a shear production range (II) and an inertial range (III).}
	\label{fig:EigenValues_convergence_scaling}
\end{figure}
\begin{table}
	\begin{tabular}{c r r l c c }
		\hline
		$n$ & $k_1/k_1^*$ & $k_2/k_2^*$& $m\quad$& $\Tstrut\Bstrut\lambda_n/u_*^2$ & Degeneracy \\
		\hline
		1-4&$\pm 1$\Tstrut & $\pm 6$ & $0$ & $0.02745$ & $4$\\
		5-6&$0$ & $\pm 6$ & $0$ & $0.02132$ & $2$ \\
		7-10&$\pm 1$ & $\pm 5$ & $0$ & $0.01190$ & $4$\\
		11-12&$0$ & $\pm 4$ & $0$ & $0.01186$ & $2$\\
		13-16&$\pm 1$ & $\pm 2$ & $0$ & $0.01074$ & $4$\\
		\hline
	\end{tabular}
	\caption{Summary of the properties of the most energetic modes $\boldsymbol{\phi}^{n}$. The wave numbers are normalized by $k_i^* = 2\pi/L_i$. The degeneracy denotes the multiplicity of the eigenvalues. }
	\label{tab:POD_large}
\end{table}

We now take a more in-depth look at the converged eigenvalue spectrum of the flow. Fig. \ref{fig:EigenValues_convergence_scaling} (a) shows the eigenvalues as a function of the index number, which is a decreasing function due to the ordering, and Fig.~\ref{fig:EigenValues_convergence_scaling} (b) shows the pre-multiplied spectrum, providing a graphical impression of the energy distribution in a log-scale plot (i.e $\sum \lambda_n \approx \int \overline{\lambda}(n) \, {\rm d}n = \int n \overline{\lambda}(n) \, {\rm d}\log(n)$, with $\overline{\lambda}(n)$ the continuous extension of $\lambda_n$, e.g. by using linear interpolation). Similar to the classical boundary layer spectrum, three different regions seem to exist, also marked on the figure. 

A first region contains the most energetic modes. Table \ref{tab:POD_large} summarizes the wave numbers, vertical model numbers $m$, and eigenvalues of the first sixteen most energetic modes. They are all very long in the streamwise direction, and have vertical mode number $0$. Note that, despite carrying the most energy per mode, they only have a modest contribution to the total energy, because they are relatively few in total numbers (see Fig. \ref{fig:EigenValues_convergence_scaling} (b) for an appraisal of distribution of energy over the modes in a log-scale plot)

A second region shows a $\lambda \sim n^{-1}$ spectrum. Although this reminds of the well known $k^{-1}$ scaling of the streamwise energy spectrum in turbulent boundary layers in the shear production range \cite{perry1986theoretical}, a formal connection has not been established to our knowledge. Finally, a third region exhibits  $\lambda \sim n^{-11/9}$ scaling. This corresponds with inertial-range scaling, as was demonstrated by Ref.~\cite{knight1990kolmogorov}, and later proven more rigorously in Ref.~\cite{moser1994kolmogorov}.

\subsection{Estimation of the effect of subgrid-scale energy} \label{ss:CGcomp}
\begin{figure}
	\centering
	\includegraphics[width=0.8\textwidth]{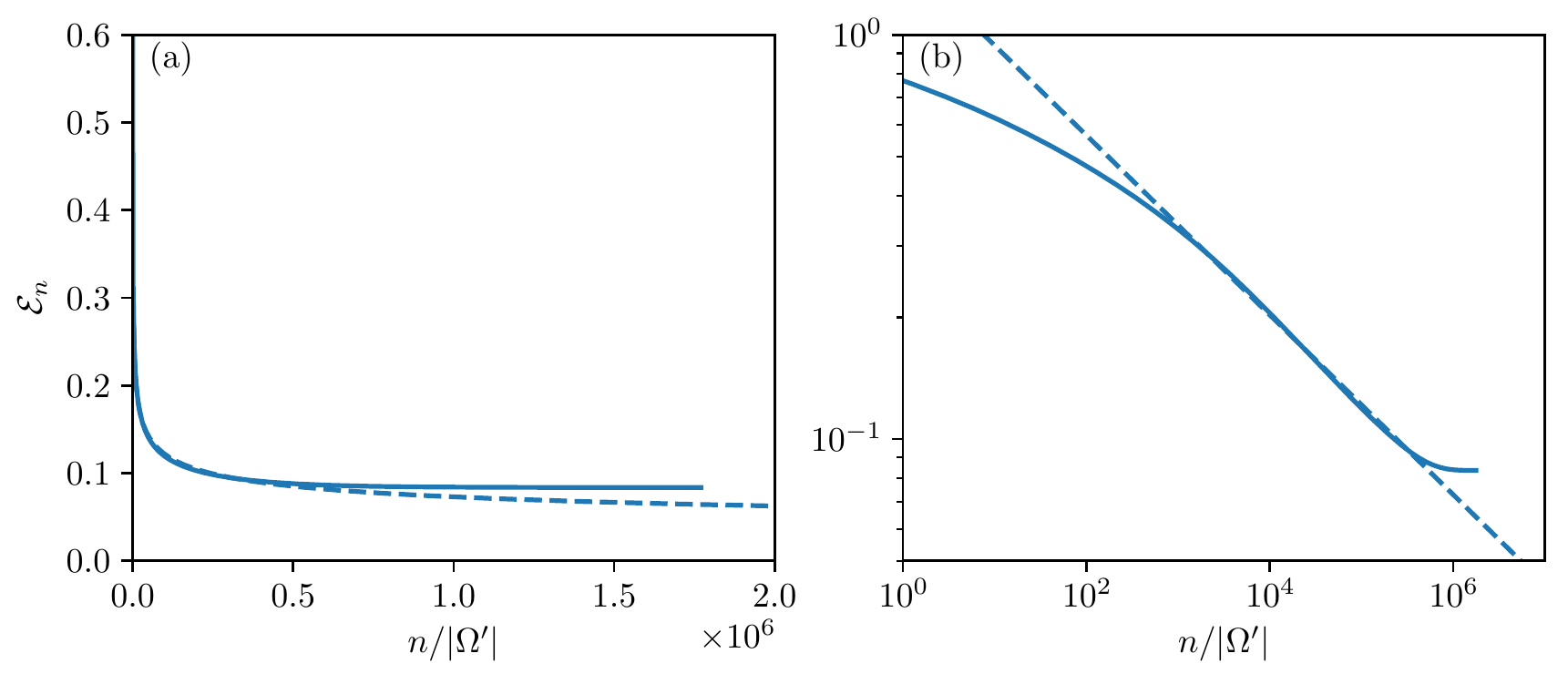}%
	\caption{Fraction of unresolved energy $\mathcal{E}_n$ as a function of the amount of modes normalized by the volume $n/|\Omega'|$. (\protect\tikz[baseline=-0.8ex]\protect\draw [blue,thick] (0,0) -- (0.8,0);) POD data, (\protect\tikz[baseline=-0.8ex]\protect\draw [blue,thick,dash pattern={on 4pt off 2pt}] (0,0) -- (0.8,0);) fitting the data to $\mathcal{E}_n\propto (n/|\Omega'|)^{-2/9}$.}
	\label{fig:energy_convergence}
\end{figure}

The slow decrease in the energy of the KL modes with increasing mode number (i.e. $\lambda \sim n^{-11/9}$ in the inertial range) suggests that the KL-dimension may be sensitive to the fraction of unresolved energy in the LES. In this section we estimate this unresolved energy based on the power law scaling found in the inertial zone. Assuming that the LES filter cut-off is in the inertial range, it is easily shown, by integrating a $n^{-11/9}$ spectrum from $n$ to $\infty$, that the residual kinetic energy $K-K_{n}$ scales as $K-K_{n}\sim n^{-2/9}$, such that the normalized residual $\mathcal{E}_{n}$ can be expressed as
\begin{equation} \label{eq:KL_conv}
\mathcal{E}_{n} = 1-\frac{K_{n}}{K} = C_{\mt{KL}}\left(\frac{n}{|\Omega'|}\right)^{-2/9},
\end{equation}
with $C_{\mt{KL}}$ and $K$ parameters that need to be further identified. Equation \ref{eq:KL_conv} is expected to hold far enough from the wall where the filter falls in the inertial range, i.e. $x_3\gg \Delta$, but is not valid close to the wall. The contribution of the near-wall region to the TKE is estimated in Appendix~\ref{s:near_wall_ener} to be on the order $1\%$, and is further neglected here.

We find the parameters $C_{\mt{KL}}$ and $K$ in Eq.~(\ref{eq:KL_conv}) by a least squares fit using the data of $K_n$ from \S \ref{ss:spectrum_KLdim}, in the range $n/|\Omega'|$ from $10^3$ to $10^5$, resulting in $C_{\mt{KL}}=1.57$ and $K= \SI{2.56}{u_*^2}$. In Fig. \ref{fig:energy_convergence} we show the result of this fitting. We find from the asymptotic behaviour of $\mathcal{E}_n$ to high $n$, that still a significant portion of the energy is unresolved in the LES, i.e  $\mathcal{E}_{\Delta}\approx 8.3\%$, where
\begin{equation} \label{eq:energy_frac_unresol}
\mathcal{E}_{\Delta}\triangleq 1-\frac{K_{\Delta}}{K}=\mathcal{E}_{N_\mt{m}}
\end{equation}
This is higher than, e.g., reported in \cite{chapman1979computational, stevens2014large} at similar simulation resolutions, and therefore we further verify this number based an an alternative method. 

We start by introducing the spectral energy tensor for isotropic turbulence
  \begin{equation}
  \Phi_{ij}(\bs{k},\bs{x}) = \frac{E(k,x_3)}{4\pi k^2}\left(\delta_{ij} -\frac{k_ik_j}{k^2}\right),
  \end{equation}
where $\bs{k}=[k_1,k_2,k_3]$ is the wave vector and $k$ its magnitude. For the energy $E(k,x_3)$ the height-dependent Kolmogorov energy spectrum $E(k,x_3)=C_{\mt{K}}\varepsilon^{2/3}k^{-5/3}$ is used, with $C_{\mt{K}}\approx1.6$ the Kolmogorov constant and $\varepsilon$ the local dissipation of turbulent kinetic energy. For the dissipation $\varepsilon$ we use the usual hypothesis that local production equals dissipation such that  $\varepsilon \approx \kappa^{-1} u_*^3(1-x_3/H)/ x_3$, with $\kappa\approx0.4$ the von Kármán constant. An estimate of the unresolved kinetic energy $K-K_{\Delta}$ is obtained by integrating the spectral energy tensor over the domain $\Omega$ and over the unresolved wave numbers. The choice of the cut-off for the unresolved wave numbers is a bit arbitrary, and for convenience we choose an isotropic equivalent cut-off wave number $k_{\Delta}=\pi/\Delta$, with $\Delta = (\Delta_1\Delta_2\Delta_3)^{1/3}$ the characteristic grid spacing. This leads to
\begin{align} \label{eq:energy_frac}
\frac{K-K_{\Delta}}{u_*^2} &= \frac{1}{|\Omega|u_*^2}\int_{\Omega}\int_{|\bs{k}|\geq k_{\Delta}} \frac{1}{2} \Phi_{ii}(\bs{k},\bs{x})\, \mt{d}\bs{k} \mt{d}\bs{x} = \frac{2\pi}{\sqrt{3}} C_{\mt{K}}\left(\kappa k_{\Delta} H\right)^{-2/3}.
\end{align}
Using $k_{\Delta}$ from our simulation, we find $K-K_{\Delta}=\SI{0.237}{u_*^2}$. Further using $K_{\Delta}=\SI{2.29}{u_*^2}$ from the LES, then leads to $K=\SI{2.52}{u_*^2}$ and $\mathcal{E}_{\Delta}=9.3\%$, which is remarkably close to the values found by the asymptotic behaviour of the grid and the POD-modes.

Finally, as a further validation, we compare the variance of the streamwise velocity component in our simulations with experimental data. Here we compare with measurements at the SLTEST site \cite{hutchins2012towards,marusic2013logarithmic}, using $u_*=\SI{0.1884}{\metre. \second^{-1}}$, $H = \SI{60}{\metre}$ from Ref. \cite{marusic2013logarithmic}. It is observed that the  LES data are consistently lower than the measurement data. Corrected LES data for the unresolved energy are also shown in the figure, based on 
\begin{equation}
\langle u_1' u_1'\rangle_{t} - \langle \ft{u}_1' \ft{u}_1'\rangle_{t}\approx\int_{|\bs{k}|\geq k_{\Delta}} \Phi_{11}(\bs{k},\bs{x}) \,\mt{d}\bs{k}= C_{\mt{K}}\varepsilon^{2/3}k_{\Delta}^{-2/3},
\end{equation}
and using the height-dependent dissipation estimate from above. These corrected LES data better fit the experiments, but it should be noted that significant uncertainty exists on the measurement data, related to the estimation of BL-height and the friction velocity. Therefore, as suggested in Ref.~\cite{hutchins2012towards}, we have added uncertainty bars on the experimental data that correspond to an error of $10\%$ on the value of $u_*$. 

\begin{figure}
	\centering
	\includegraphics[width=0.5\textwidth]{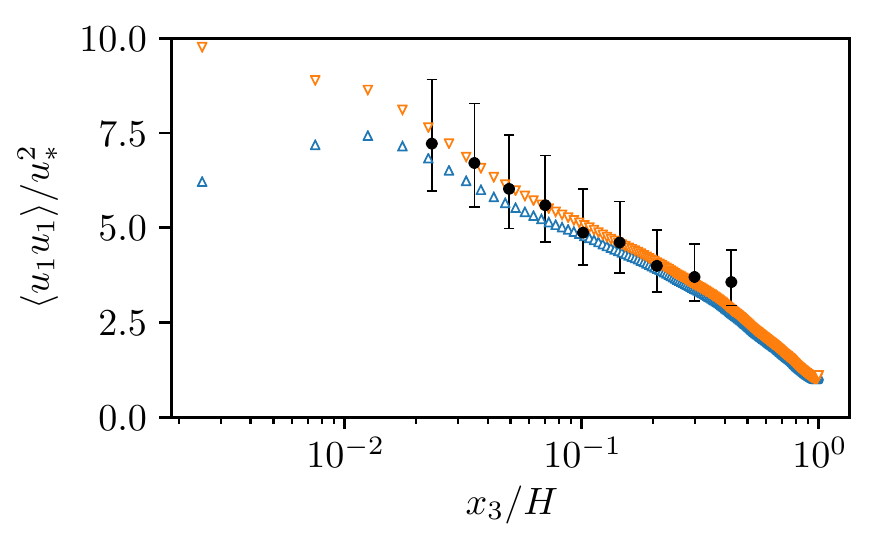}%
	\caption{Variance of streamwise velocity component. (\protect\tikz{\protect\filldraw[draw=blue,fill=white] (0,0.0cm) --
			(0.2cm,0.0cm) -- (0.1cm,0.2cm) -- (0,0.0cm);}) represents $\langle \ft{u}_1'\ft{u}_1' \rangle$, (\protect\tikz{\protect\filldraw[draw=orange,fill=white] (0,0.2cm) --
			(0.2cm,0.2cm) -- (0.1cm,0.0cm) -- (0,0.2cm);}) represents a correction for the unresolved energy $\langle u_1' u_1'\rangle $, (\protect\tikz{\protect\filldraw[draw=black,fill=black] (0,0) circle [radius=0.1cm];}) neutral ABL measurement data from Ref. \cite{marusic2013logarithmic}. The error bars indicate the $10\%$ uncertainty intervals on the friction velocity $u_*$.}
	\label{fig:BL_inst}
\end{figure}

In summary, we have found that the fraction of unresolved energy of the reference simulation is estimated at $\mathcal{E}_{\Delta}=8.4\%$, and therefore, the KL dimension $d_{\mt{KL}}^{\Delta}$ obtained in \S\ref{ss:spectrum_KLdim} is an underestimation of the true value $d_{\mt{KL}}$. A better estimate of the KL dimension is then obtained by inverting Eq.~\ref{eq:KL_conv}, which leads to $n/|\Omega'| = (\mathcal{E}_n/C_{\mt{KL}})^{-9/2}$. This yields an expression for the number of modes required to express a specified unresolved fraction of energy $\mathcal{E}_n$. Using $\mathcal{E}_n = 0.1$ gives a KL-dimension $d_{\mt{KL}}=2.41\times10^5$, which is more than a factor 10 larger than the initial estimate determined earlier in \S\ref{ss:convergence_KL_dimension}.

\section{Conclusions}
We performed a Karhunen--Loève decomposition for a LES of a high Reynolds number pressure-driven boundary layer. We conclude that to resolve $90\%$ of the TKE on average -- the so called KL-dimension -- $\SI{2.4e5}{|\Omega|/H^3}$ modes are needed, which is up to three orders of magnitude higher than values commonly reported in earlier studies. This indicates that more caution should be exercised when considering the convergence of POD basis in high-Reynolds number boundary layers such as the planetary boundary layer, and illustrates once more the challenges associated to representing turbulence in a low-dimensional basis.

\appendix
\section{Estimation of the near wall energy} \label{s:near_wall_ener}
The unresolved energy close to the wall is estimated by making a rough estimate of the integral
\begin{equation}
\frac{1}{u_*^2H}\int_{0}^{\Delta} \langle u_i' u_i' \rangle_t\,\mt{d}x_3.
\end{equation}
We proceed in two steps, first the energy below the logarithmic region is estimated (i.e. the roughness sublayer), and secondly the contribution of the logarithmic region. The roughness sublayer is very narrow compared to the BL height ($z_0/H\ll 1$), and although there is a peak of turbulent kinetic energy, its total contribution is therefore negligible. Similar considerations hold for smooth walls, for which, e.g., the peak of TKE scales with $\sim u_*^2\log \mt{Re}$  for smooth walls (see e.g. \cite{meneveau2013generalized}) and becomes lower with increasing wall roughness \cite{jimenez2004turbulent}, while the width below the log region scales with  $Re_{\tau}^{-1/2}H$ \cite{marusic2013logarithmic}, such that the fraction of energy in this region scales at most with
\begin{equation}
\frac{1}{H}\int_{0}^{Re_{\tau}^{-1/2}H}\log \mt{Re}\,\mt{d}x_3= Re_{\tau}^{-1/2}\log Re_{\tau},
\end{equation}
which equals e.g. $5\times10^{-3}$ for $Re_{\tau}=10^7$ a typical value in the atmospheric boundary layer. To estimate the contribution of the unresolved energy in the logarithmic region, we employ Townsend's similarity hypothesis for the velocity fluctuation \cite{townsend1976structure}, i.e.  $\langle u_i'u_i' \rangle_t/u_*^2=B - A\log(x_3/H)$, usually expressed per velocity component $\langle u_i^{\prime2} \rangle/u_*^2=B_i - A_i\log(x_3/H)$, with $B=B_1+B_2+B_3$ and $A=A_1+A_2$. Integrating from $0$ to $\Delta$ and normalizing by $u_*^2H$ leads to an estimate of the energy in this region
\begin{equation}
\frac{1}{H}\int_{0}^{\Delta}B - A\log(x_3/H)\,\mt{d}x_3= \frac{\Delta}{H} \left(A+B-B\log\left(\frac{\Delta}{H}\right)\right),
\end{equation}
which is typically $\mathcal{O}(10^{-2})$ and therefore contributions to $\mathcal{E}_{n}$ are expected to be of similar magnitude, and are therefore negligible.


\begin{thebibliography}{57}%
	\makeatletter
	\providecommand \@ifxundefined [1]{%
		\@ifx{#1\undefined}
	}%
	\providecommand \@ifnum [1]{%
		\ifnum #1\expandafter \@firstoftwo
		\else \expandafter \@secondoftwo
		\fi
	}%
	\providecommand \@ifx [1]{%
		\ifx #1\expandafter \@firstoftwo
		\else \expandafter \@secondoftwo
		\fi
	}%
	\providecommand \natexlab [1]{#1}%
	\providecommand \enquote  [1]{``#1''}%
	\providecommand \bibnamefont  [1]{#1}%
	\providecommand \bibfnamefont [1]{#1}%
	\providecommand \citenamefont [1]{#1}%
	\providecommand \href@noop [0]{\@secondoftwo}%
	\providecommand \href [0]{\begingroup \@sanitize@url \@href}%
	\providecommand \@href[1]{\@@startlink{#1}\@@href}%
	\providecommand \@@href[1]{\endgroup#1\@@endlink}%
	\providecommand \@sanitize@url [0]{\catcode `\\12\catcode `\$12\catcode
		`\&12\catcode `\#12\catcode `\^12\catcode `\_12\catcode `\%12\relax}%
	\providecommand \@@startlink[1]{}%
	\providecommand \@@endlink[0]{}%
	\providecommand \url  [0]{\begingroup\@sanitize@url \@url }%
	\providecommand \@url [1]{\endgroup\@href {#1}{\urlprefix }}%
	\providecommand \urlprefix  [0]{URL }%
	\providecommand \Eprint [0]{\href }%
	\providecommand \doibase [0]{https://doi.org/}%
	\providecommand \selectlanguage [0]{\@gobble}%
	\providecommand \bibinfo  [0]{\@secondoftwo}%
	\providecommand \bibfield  [0]{\@secondoftwo}%
	\providecommand \translation [1]{[#1]}%
	\providecommand \BibitemOpen [0]{}%
	\providecommand \bibitemStop [0]{}%
	\providecommand \bibitemNoStop [0]{.\EOS\space}%
	\providecommand \EOS [0]{\spacefactor3000\relax}%
	\providecommand \BibitemShut  [1]{\csname bibitem#1\endcsname}%
	\let\auto@bib@innerbib\@empty
	\bibitem [{\citenamefont {Kaplan}\ and\ \citenamefont
		{Yorke}(1979)}]{kaplan1979chaotic}%
	\BibitemOpen
	\bibfield  {author} {\bibinfo {author} {\bibfnamefont {J.~L.}\ \bibnamefont
			{Kaplan}}\ and\ \bibinfo {author} {\bibfnamefont {J.~A.}\ \bibnamefont
			{Yorke}},\ }\bibfield  {title} {\bibinfo {title} {Chaotic behavior of
			multidimensional difference equations},\ }in\ \href@noop {} {\emph {\bibinfo
			{booktitle} {Functional differential equations and approximation of fixed
				points}}}\ (\bibinfo  {publisher} {Springer},\ \bibinfo {year} {1979})\ pp.\
	\bibinfo {pages} {204--227}\BibitemShut {NoStop}%
	\bibitem [{\citenamefont {Pearson}(1901)}]{pearson1901liii}%
	\BibitemOpen
	\bibfield  {author} {\bibinfo {author} {\bibfnamefont {K.}~\bibnamefont
			{Pearson}},\ }\bibfield  {title} {\bibinfo {title} {{LIII.} {O}n lines and
			planes of closest fit to systems of points in space},\ }\href@noop {}
	{\bibfield  {journal} {\bibinfo  {journal} {The London, Edinburgh, and Dublin
				Philosophical Magazine and Journal of Science}\ }\textbf {\bibinfo {volume}
			{2}},\ \bibinfo {pages} {559} (\bibinfo {year} {1901})}\BibitemShut {NoStop}%
	\bibitem [{\citenamefont {Sirovich}(1989)}]{sirovich1989chaotic}%
	\BibitemOpen
	\bibfield  {author} {\bibinfo {author} {\bibfnamefont {L.}~\bibnamefont
			{Sirovich}},\ }\bibfield  {title} {\bibinfo {title} {Chaotic dynamics of
			coherent structures},\ }\href@noop {} {\bibfield  {journal} {\bibinfo
			{journal} {Physica D: Nonlin. Phen.}\ }\textbf {\bibinfo {volume} {37}},\
		\bibinfo {pages} {126} (\bibinfo {year} {1989})}\BibitemShut {NoStop}%
	\bibitem [{\citenamefont {Sirovich}\ and\ \citenamefont
		{Deane}(1991)}]{sirovich1991computational}%
	\BibitemOpen
	\bibfield  {author} {\bibinfo {author} {\bibfnamefont {L.}~\bibnamefont
			{Sirovich}}\ and\ \bibinfo {author} {\bibfnamefont {A.~E.}\ \bibnamefont
			{Deane}},\ }\bibfield  {title} {\bibinfo {title} {A computational study of
			{R}ayleigh--{B}{\'e}nard convection. {P}art 2. {D}imension considerations},\
	}\href@noop {} {\bibfield  {journal} {\bibinfo  {journal} {J. Fluid Mech.}\
		}\textbf {\bibinfo {volume} {222}},\ \bibinfo {pages} {251} (\bibinfo {year}
		{1991})}\BibitemShut {NoStop}%
	\bibitem [{\citenamefont {Berkooz}\ \emph {et~al.}(1993)\citenamefont
		{Berkooz}, \citenamefont {Holmes},\ and\ \citenamefont
		{Lumley}}]{berkooz1993proper}%
	\BibitemOpen
	\bibfield  {author} {\bibinfo {author} {\bibfnamefont {G.}~\bibnamefont
			{Berkooz}}, \bibinfo {author} {\bibfnamefont {P.}~\bibnamefont {Holmes}},\
		and\ \bibinfo {author} {\bibfnamefont {J.~L.}\ \bibnamefont {Lumley}},\
	}\bibfield  {title} {\bibinfo {title} {The proper orthogonal decomposition in
			the analysis of turbulent flows},\ }\href@noop {} {\bibfield  {journal}
		{\bibinfo  {journal} {Ann. Rev. Fluid Mech.}\ }\textbf {\bibinfo {volume}
			{25}},\ \bibinfo {pages} {539} (\bibinfo {year} {1993})}\BibitemShut
	{NoStop}%
	\bibitem [{\citenamefont
		{Sirovich}(1987{\natexlab{a}})}]{sirovich1987turbulence}%
	\BibitemOpen
	\bibfield  {author} {\bibinfo {author} {\bibfnamefont {L.}~\bibnamefont
			{Sirovich}},\ }\bibfield  {title} {\bibinfo {title} {Turbulence and the
			dynamics of coherent structures. {I. C}oherent structures},\ }\href@noop {}
	{\bibfield  {journal} {\bibinfo  {journal} {Q. Appl. Math}\ }\textbf
		{\bibinfo {volume} {45}},\ \bibinfo {pages} {561} (\bibinfo {year}
		{1987}{\natexlab{a}})}\BibitemShut {NoStop}%
	\bibitem [{\citenamefont {Duggleby}\ \emph {et~al.}(2008)\citenamefont
		{Duggleby}, \citenamefont {Ball},\ and\ \citenamefont
		{Schwaenen}}]{duggleby2008structure}%
	\BibitemOpen
	\bibfield  {author} {\bibinfo {author} {\bibfnamefont {A.}~\bibnamefont
			{Duggleby}}, \bibinfo {author} {\bibfnamefont {K.~S.}\ \bibnamefont {Ball}},\
		and\ \bibinfo {author} {\bibfnamefont {M.}~\bibnamefont {Schwaenen}},\
	}\bibfield  {title} {\bibinfo {title} {Structure and dynamics of low
			{R}eynolds number turbulent pipe flow},\ }\href@noop {} {\bibfield  {journal}
		{\bibinfo  {journal} {Phil. Trans. R. Soc. A}\ }\textbf {\bibinfo {volume}
			{367}},\ \bibinfo {pages} {473} (\bibinfo {year} {2008})}\BibitemShut
	{NoStop}%
	\bibitem [{\citenamefont {Ball}\ \emph {et~al.}(1991)\citenamefont {Ball},
		\citenamefont {Sirovich},\ and\ \citenamefont {Keefe}}]{ball1991dynamical}%
	\BibitemOpen
	\bibfield  {author} {\bibinfo {author} {\bibfnamefont {K.}~\bibnamefont
			{Ball}}, \bibinfo {author} {\bibfnamefont {L.}~\bibnamefont {Sirovich}},\
		and\ \bibinfo {author} {\bibfnamefont {L.}~\bibnamefont {Keefe}},\ }\bibfield
	{title} {\bibinfo {title} {Dynamical eigenfunction decomposition of
			turbulent channel flow},\ }\href@noop {} {\bibfield  {journal} {\bibinfo
			{journal} {Int. J. Numer. Methods Fluids}\ }\textbf {\bibinfo {volume}
			{12}},\ \bibinfo {pages} {585} (\bibinfo {year} {1991})}\BibitemShut
	{NoStop}%
	\bibitem [{\citenamefont {Webber}\ \emph {et~al.}(1997)\citenamefont {Webber},
		\citenamefont {Handler},\ and\ \citenamefont
		{Sirovich}}]{webber1997karhunen}%
	\BibitemOpen
	\bibfield  {author} {\bibinfo {author} {\bibfnamefont {G.~A.}\ \bibnamefont
			{Webber}}, \bibinfo {author} {\bibfnamefont {R.}~\bibnamefont {Handler}},\
		and\ \bibinfo {author} {\bibfnamefont {L.}~\bibnamefont {Sirovich}},\
	}\bibfield  {title} {\bibinfo {title} {The {K}arhunen--{L}oeve decomposition
			of minimal channel flow},\ }\href@noop {} {\bibfield  {journal} {\bibinfo
			{journal} {Phys. Fluids}\ }\textbf {\bibinfo {volume} {9}},\ \bibinfo {pages}
		{1054} (\bibinfo {year} {1997})}\BibitemShut {NoStop}%
	\bibitem [{\citenamefont {Sirovich}\ \emph {et~al.}(1991)\citenamefont
		{Sirovich}, \citenamefont {Ball},\ and\ \citenamefont
		{Handler}}]{sirovich1991propagating}%
	\BibitemOpen
	\bibfield  {author} {\bibinfo {author} {\bibfnamefont {L.}~\bibnamefont
			{Sirovich}}, \bibinfo {author} {\bibfnamefont {K.}~\bibnamefont {Ball}},\
		and\ \bibinfo {author} {\bibfnamefont {R.}~\bibnamefont {Handler}},\
	}\bibfield  {title} {\bibinfo {title} {Propagating structures in wall-bounded
			turbulent flows},\ }\href@noop {} {\bibfield  {journal} {\bibinfo  {journal}
			{Theor. Comp. Fluid Dyn.}\ }\textbf {\bibinfo {volume} {2}},\ \bibinfo
		{pages} {307} (\bibinfo {year} {1991})}\BibitemShut {NoStop}%
	\bibitem [{\citenamefont {Iwamoto}\ \emph {et~al.}(2002)\citenamefont
		{Iwamoto}, \citenamefont {Suzuki},\ and\ \citenamefont
		{Kasagi}}]{iwamoto2002reynolds}%
	\BibitemOpen
	\bibfield  {author} {\bibinfo {author} {\bibfnamefont {K.}~\bibnamefont
			{Iwamoto}}, \bibinfo {author} {\bibfnamefont {Y.}~\bibnamefont {Suzuki}},\
		and\ \bibinfo {author} {\bibfnamefont {N.}~\bibnamefont {Kasagi}},\
	}\bibfield  {title} {\bibinfo {title} {Reynolds number effect on wall
			turbulence: toward effective feedback control},\ }\href@noop {} {\bibfield
		{journal} {\bibinfo  {journal} {Int J. Heat Fluid Flow}\ }\textbf {\bibinfo
			{volume} {23}},\ \bibinfo {pages} {678} (\bibinfo {year} {2002})}\BibitemShut
	{NoStop}%
	\bibitem [{\citenamefont {Housiadas}\ \emph {et~al.}(2005)\citenamefont
		{Housiadas}, \citenamefont {Beris},\ and\ \citenamefont
		{Handler}}]{housiadas2005viscoelastic}%
	\BibitemOpen
	\bibfield  {author} {\bibinfo {author} {\bibfnamefont {K.~D.}\ \bibnamefont
			{Housiadas}}, \bibinfo {author} {\bibfnamefont {A.~N.}\ \bibnamefont
			{Beris}},\ and\ \bibinfo {author} {\bibfnamefont {R.~A.}\ \bibnamefont
			{Handler}},\ }\bibfield  {title} {\bibinfo {title} {Viscoelastic effects on
			higher order statistics and on coherent structures in turbulent channel
			flow},\ }\href@noop {} {\bibfield  {journal} {\bibinfo  {journal} {Phys.
				Fluids}\ }\textbf {\bibinfo {volume} {17}},\ \bibinfo {pages} {035106}
		(\bibinfo {year} {2005})}\BibitemShut {NoStop}%
	\bibitem [{\citenamefont {Smith}\ \emph {et~al.}(2005)\citenamefont {Smith},
		\citenamefont {Moehlis},\ and\ \citenamefont {Holmes}}]{smith2005low}%
	\BibitemOpen
	\bibfield  {author} {\bibinfo {author} {\bibfnamefont {T.}~\bibnamefont
			{Smith}}, \bibinfo {author} {\bibfnamefont {J.}~\bibnamefont {Moehlis}},\
		and\ \bibinfo {author} {\bibfnamefont {P.}~\bibnamefont {Holmes}},\
	}\bibfield  {title} {\bibinfo {title} {Low-dimensional models for turbulent
			plane {C}ouette flow in a minimal flow unit},\ }\href@noop {} {\bibfield
		{journal} {\bibinfo  {journal} {J. Fluid Mech.}\ }\textbf {\bibinfo {volume}
			{538}},\ \bibinfo {pages} {71} (\bibinfo {year} {2005})}\BibitemShut
	{NoStop}%
	\bibitem [{\citenamefont {Ciliberto}\ and\ \citenamefont
		{Nicolaenko}(1991)}]{ciliberto1991estimating}%
	\BibitemOpen
	\bibfield  {author} {\bibinfo {author} {\bibfnamefont {S.}~\bibnamefont
			{Ciliberto}}\ and\ \bibinfo {author} {\bibfnamefont {B.}~\bibnamefont
			{Nicolaenko}},\ }\bibfield  {title} {\bibinfo {title} {Estimating the number
			of degrees of freedom in spatially extended systems},\ }\href@noop {}
	{\bibfield  {journal} {\bibinfo  {journal} {Europhys. Lett.}\ }\textbf
		{\bibinfo {volume} {14}},\ \bibinfo {pages} {303} (\bibinfo {year}
		{1991})}\BibitemShut {NoStop}%
	\bibitem [{\citenamefont {Duggleby}\ \emph {et~al.}(2007)\citenamefont
		{Duggleby}, \citenamefont {Ball}, \citenamefont {Paul},\ and\ \citenamefont
		{Fischer}}]{duggleby2007dynamical}%
	\BibitemOpen
	\bibfield  {author} {\bibinfo {author} {\bibfnamefont {A.}~\bibnamefont
			{Duggleby}}, \bibinfo {author} {\bibfnamefont {K.~S.}\ \bibnamefont {Ball}},
		\bibinfo {author} {\bibfnamefont {M.~R.}\ \bibnamefont {Paul}},\ and\
		\bibinfo {author} {\bibfnamefont {P.~F.}\ \bibnamefont {Fischer}},\
	}\bibfield  {title} {\bibinfo {title} {Dynamical eigenfunction decomposition
			of turbulent pipe flow},\ }\href@noop {} {\bibfield  {journal} {\bibinfo
			{journal} {J. Turbul.}\ ,\ \bibinfo {pages} {N43}} (\bibinfo {year}
		{2007})}\BibitemShut {NoStop}%
	\bibitem [{\citenamefont {Cardillo}\ \emph {et~al.}(2013)\citenamefont
		{Cardillo}, \citenamefont {Chen}, \citenamefont {Araya}, \citenamefont
		{Newman}, \citenamefont {Jansen},\ and\ \citenamefont
		{Castillo}}]{cardillo2013dns}%
	\BibitemOpen
	\bibfield  {author} {\bibinfo {author} {\bibfnamefont {J.}~\bibnamefont
			{Cardillo}}, \bibinfo {author} {\bibfnamefont {Y.}~\bibnamefont {Chen}},
		\bibinfo {author} {\bibfnamefont {G.}~\bibnamefont {Araya}}, \bibinfo
		{author} {\bibfnamefont {J.}~\bibnamefont {Newman}}, \bibinfo {author}
		{\bibfnamefont {K.}~\bibnamefont {Jansen}},\ and\ \bibinfo {author}
		{\bibfnamefont {L.}~\bibnamefont {Castillo}},\ }\bibfield  {title} {\bibinfo
		{title} {{DNS} of a turbulent boundary layer with surface roughness},\
	}\href@noop {} {\bibfield  {journal} {\bibinfo  {journal} {J. Fluid Mech.}\
		}\textbf {\bibinfo {volume} {729}},\ \bibinfo {pages} {603} (\bibinfo {year}
		{2013})}\BibitemShut {NoStop}%
	\bibitem [{\citenamefont {VerHulst}\ and\ \citenamefont
		{Meneveau}(2014)}]{verhulst2014large}%
	\BibitemOpen
	\bibfield  {author} {\bibinfo {author} {\bibfnamefont {C.}~\bibnamefont
			{VerHulst}}\ and\ \bibinfo {author} {\bibfnamefont {C.}~\bibnamefont
			{Meneveau}},\ }\bibfield  {title} {\bibinfo {title} {Large eddy simulation
			study of the kinetic energy entrainment by energetic turbulent flow
			structures in large wind farms},\ }\href@noop {} {\bibfield  {journal}
		{\bibinfo  {journal} {Phys. Fluids}\ }\textbf {\bibinfo {volume} {26}},\
		\bibinfo {pages} {025113} (\bibinfo {year} {2014})}\BibitemShut {NoStop}%
	\bibitem [{\citenamefont {Ali}\ \emph {et~al.}(2017)\citenamefont {Ali},
		\citenamefont {Cortina}, \citenamefont {Hamilton}, \citenamefont {Calaf},\
		and\ \citenamefont {Cal}}]{ali2017turbulence}%
	\BibitemOpen
	\bibfield  {author} {\bibinfo {author} {\bibfnamefont {N.}~\bibnamefont
			{Ali}}, \bibinfo {author} {\bibfnamefont {G.}~\bibnamefont {Cortina}},
		\bibinfo {author} {\bibfnamefont {N.}~\bibnamefont {Hamilton}}, \bibinfo
		{author} {\bibfnamefont {M.}~\bibnamefont {Calaf}},\ and\ \bibinfo {author}
		{\bibfnamefont {R.}~\bibnamefont {Cal}},\ }\bibfield  {title} {\bibinfo
		{title} {Turbulence characteristics of a thermally stratified wind turbine
			array boundary layer via proper orthogonal decomposition},\ }\href@noop {}
	{\bibfield  {journal} {\bibinfo  {journal} {J. Fluid Mech.}\ }\textbf
		{\bibinfo {volume} {828}},\ \bibinfo {pages} {175} (\bibinfo {year}
		{2017})}\BibitemShut {NoStop}%
	\bibitem [{\citenamefont {Zhang}\ and\ \citenamefont
		{Stevens}(2019)}]{zhang2019characterizing}%
	\BibitemOpen
	\bibfield  {author} {\bibinfo {author} {\bibfnamefont {M.}~\bibnamefont
			{Zhang}}\ and\ \bibinfo {author} {\bibfnamefont {R.~J.}\ \bibnamefont
			{Stevens}},\ }\bibfield  {title} {\bibinfo {title} {Characterizing the
			coherent structures within and above large wind farms},\ }\href@noop {}
	{\bibfield  {journal} {\bibinfo  {journal} {Boundary-Layer Meteorol.}\ ,\
			\bibinfo {pages} {1}} (\bibinfo {year} {2019})}\BibitemShut {NoStop}%
	\bibitem [{\citenamefont {Shah}\ and\ \citenamefont
		{Bou-Zeid}(2014)}]{shah2014very}%
	\BibitemOpen
	\bibfield  {author} {\bibinfo {author} {\bibfnamefont {S.}~\bibnamefont
			{Shah}}\ and\ \bibinfo {author} {\bibfnamefont {E.}~\bibnamefont
			{Bou-Zeid}},\ }\bibfield  {title} {\bibinfo {title} {Very-large-scale motions
			in the atmospheric boundary layer educed by snapshot proper orthogonal
			decomposition},\ }\href@noop {} {\bibfield  {journal} {\bibinfo  {journal}
			{Boundary-Layer Meteorol.}\ }\textbf {\bibinfo {volume} {153}},\ \bibinfo
		{pages} {355} (\bibinfo {year} {2014})}\BibitemShut {NoStop}%
	\bibitem [{\citenamefont {Andersen}\ \emph {et~al.}(2014)\citenamefont
		{Andersen}, \citenamefont {S{\o}rensen},\ and\ \citenamefont
		{Mikkelsen}}]{andersen2014reduced}%
	\BibitemOpen
	\bibfield  {author} {\bibinfo {author} {\bibfnamefont {S.~J.}\ \bibnamefont
			{Andersen}}, \bibinfo {author} {\bibfnamefont {J.~N.}\ \bibnamefont
			{S{\o}rensen}},\ and\ \bibinfo {author} {\bibfnamefont {R.}~\bibnamefont
			{Mikkelsen}},\ }\bibfield  {title} {\bibinfo {title} {Reduced order model of
			the inherent turbulence of wind turbine wakes inside an infinitely long row
			of turbines},\ }in\ \href@noop {} {\emph {\bibinfo {booktitle} {Journal of
				Physics: Conference Series}}},\ Vol.\ \bibinfo {volume} {555}\ (\bibinfo
	{organization} {IOP Publishing},\ \bibinfo {year} {2014})\ p.\ \bibinfo
	{pages} {012005}\BibitemShut {NoStop}%
	\bibitem [{\citenamefont {Andersen}\ \emph {et~al.}(2017)\citenamefont
		{Andersen}, \citenamefont {S{\o}rensen},\ and\ \citenamefont
		{Mikkelsen}}]{andersen2017turbulence}%
	\BibitemOpen
	\bibfield  {author} {\bibinfo {author} {\bibfnamefont {S.~J.}\ \bibnamefont
			{Andersen}}, \bibinfo {author} {\bibfnamefont {J.~N.}\ \bibnamefont
			{S{\o}rensen}},\ and\ \bibinfo {author} {\bibfnamefont {R.~F.}\ \bibnamefont
			{Mikkelsen}},\ }\bibfield  {title} {\bibinfo {title} {Turbulence and
			entrainment length scales in large wind farms},\ }\href@noop {} {\bibfield
		{journal} {\bibinfo  {journal} {Phil. Trans. R. Soc. A}\ }\textbf {\bibinfo
			{volume} {375}},\ \bibinfo {pages} {20160107} (\bibinfo {year}
		{2017})}\BibitemShut {NoStop}%
	\bibitem [{\citenamefont {Bastine}\ \emph {et~al.}(2015)\citenamefont
		{Bastine}, \citenamefont {Witha}, \citenamefont {W{\"a}chter},\ and\
		\citenamefont {Peinke}}]{bastine2015towards}%
	\BibitemOpen
	\bibfield  {author} {\bibinfo {author} {\bibfnamefont {D.}~\bibnamefont
			{Bastine}}, \bibinfo {author} {\bibfnamefont {B.}~\bibnamefont {Witha}},
		\bibinfo {author} {\bibfnamefont {M.}~\bibnamefont {W{\"a}chter}},\ and\
		\bibinfo {author} {\bibfnamefont {J.}~\bibnamefont {Peinke}},\ }\bibfield
	{title} {\bibinfo {title} {Towards a simplified dynamic wake model using
			{POD} analysis},\ }\href@noop {} {\bibfield  {journal} {\bibinfo  {journal}
			{Energies}\ }\textbf {\bibinfo {volume} {8}},\ \bibinfo {pages} {895}
		(\bibinfo {year} {2015})}\BibitemShut {NoStop}%
	\bibitem [{\citenamefont {Bastankhah}\ and\ \citenamefont
		{Port{\'e}-Agel}(2017)}]{bastankhah2017wind}%
	\BibitemOpen
	\bibfield  {author} {\bibinfo {author} {\bibfnamefont {M.}~\bibnamefont
			{Bastankhah}}\ and\ \bibinfo {author} {\bibfnamefont {F.}~\bibnamefont
			{Port{\'e}-Agel}},\ }\bibfield  {title} {\bibinfo {title} {Wind tunnel study
			of the wind turbine interaction with a boundary-layer flow: Upwind region,
			turbine performance, and wake region},\ }\href@noop {} {\bibfield  {journal}
		{\bibinfo  {journal} {Physics of Fluids}\ }\textbf {\bibinfo {volume} {29}},\
		\bibinfo {pages} {065105} (\bibinfo {year} {2017})}\BibitemShut {NoStop}%
	\bibitem [{\citenamefont {Bastine}\ \emph {et~al.}(2018)\citenamefont
		{Bastine}, \citenamefont {Vollmer}, \citenamefont {W{\"a}chter},\ and\
		\citenamefont {Peinke}}]{bastine2018stochastic}%
	\BibitemOpen
	\bibfield  {author} {\bibinfo {author} {\bibfnamefont {D.}~\bibnamefont
			{Bastine}}, \bibinfo {author} {\bibfnamefont {L.}~\bibnamefont {Vollmer}},
		\bibinfo {author} {\bibfnamefont {M.}~\bibnamefont {W{\"a}chter}},\ and\
		\bibinfo {author} {\bibfnamefont {J.}~\bibnamefont {Peinke}},\ }\bibfield
	{title} {\bibinfo {title} {Stochastic wake modelling based on {POD}
			analysis},\ }\href@noop {} {\bibfield  {journal} {\bibinfo  {journal}
			{Energies}\ }\textbf {\bibinfo {volume} {11}},\ \bibinfo {pages} {612}
		(\bibinfo {year} {2018})}\BibitemShut {NoStop}%
	\bibitem [{\citenamefont {Huang}\ \emph
		{et~al.}(2009{\natexlab{a}})\citenamefont {Huang}, \citenamefont {Cassiani},\
		and\ \citenamefont {Albertson}}]{huang2009analysis}%
	\BibitemOpen
	\bibfield  {author} {\bibinfo {author} {\bibfnamefont {J.}~\bibnamefont
			{Huang}}, \bibinfo {author} {\bibfnamefont {M.}~\bibnamefont {Cassiani}},\
		and\ \bibinfo {author} {\bibfnamefont {J.}~\bibnamefont {Albertson}},\
	}\bibfield  {title} {\bibinfo {title} {Analysis of coherent structures within
			the atmospheric boundary layer},\ }\href@noop {} {\bibfield  {journal}
		{\bibinfo  {journal} {Boundary-Layer Meteorol.}\ }\textbf {\bibinfo {volume}
			{131}},\ \bibinfo {pages} {147} (\bibinfo {year}
		{2009}{\natexlab{a}})}\BibitemShut {NoStop}%
	\bibitem [{\citenamefont {Esau}(2003)}]{esau2003coriolis}%
	\BibitemOpen
	\bibfield  {author} {\bibinfo {author} {\bibfnamefont {I.~N.}\ \bibnamefont
			{Esau}},\ }\bibfield  {title} {\bibinfo {title} {The {C}oriolis effect on
			coherent structures in planetary boundary layers},\ }\href@noop {} {\bibfield
		{journal} {\bibinfo  {journal} {J. Turbul.}\ }\textbf {\bibinfo {volume}
			{4}},\ \bibinfo {pages} {1} (\bibinfo {year} {2003})}\BibitemShut {NoStop}%
	\bibitem [{\citenamefont {Keith~Wilson}(1996)}]{keith1996empirical}%
	\BibitemOpen
	\bibfield  {author} {\bibinfo {author} {\bibfnamefont {D.}~\bibnamefont
			{Keith~Wilson}},\ }\bibfield  {title} {\bibinfo {title} {Empirical orthogonal
			function analysis of the weakly convective atmospheric boundary layer. part
			{I:} eddy structures},\ }\href@noop {} {\bibfield  {journal} {\bibinfo
			{journal} {J. Atmos. Sci.}\ }\textbf {\bibinfo {volume} {53}},\ \bibinfo
		{pages} {801} (\bibinfo {year} {1996})}\BibitemShut {NoStop}%
	\bibitem [{\citenamefont {Finnigan}\ and\ \citenamefont
		{Shaw}(2000)}]{finnigan2000wind}%
	\BibitemOpen
	\bibfield  {author} {\bibinfo {author} {\bibfnamefont {J.}~\bibnamefont
			{Finnigan}}\ and\ \bibinfo {author} {\bibfnamefont {R.}~\bibnamefont
			{Shaw}},\ }\bibfield  {title} {\bibinfo {title} {A wind-tunnel study of
			airflow in waving wheat: an {EOF} analysis of the structure of the large-eddy
			motion},\ }\href@noop {} {\bibfield  {journal} {\bibinfo  {journal}
			{Boundary-Layer Meteorol.}\ }\textbf {\bibinfo {volume} {96}},\ \bibinfo
		{pages} {211} (\bibinfo {year} {2000})}\BibitemShut {NoStop}%
	\bibitem [{\citenamefont {Hamilton}\ \emph {et~al.}(2015)\citenamefont
		{Hamilton}, \citenamefont {Tutkun},\ and\ \citenamefont
		{Cal}}]{hamilton2015wind}%
	\BibitemOpen
	\bibfield  {author} {\bibinfo {author} {\bibfnamefont {N.}~\bibnamefont
			{Hamilton}}, \bibinfo {author} {\bibfnamefont {M.}~\bibnamefont {Tutkun}},\
		and\ \bibinfo {author} {\bibfnamefont {R.~B.}\ \bibnamefont {Cal}},\
	}\bibfield  {title} {\bibinfo {title} {Wind turbine boundary layer arrays for
			{Cartesian} and staggered configurations: {Part II}, low-dimensional
			representations via the proper orthogonal decomposition},\ }\href@noop {}
	{\bibfield  {journal} {\bibinfo  {journal} {Wind Energy}\ }\textbf {\bibinfo
			{volume} {18}},\ \bibinfo {pages} {297} (\bibinfo {year} {2015})}\BibitemShut
	{NoStop}%
	\bibitem [{\citenamefont {Shaw}\ and\ \citenamefont
		{Schumann}(1992)}]{shaw1992large}%
	\BibitemOpen
	\bibfield  {author} {\bibinfo {author} {\bibfnamefont {R.~H.}\ \bibnamefont
			{Shaw}}\ and\ \bibinfo {author} {\bibfnamefont {U.}~\bibnamefont
			{Schumann}},\ }\bibfield  {title} {\bibinfo {title} {Large-eddy simulation of
			turbulent flow above and within a forest},\ }\href@noop {} {\bibfield
		{journal} {\bibinfo  {journal} {Boundary-Layer Meteorology}\ }\textbf
		{\bibinfo {volume} {61}},\ \bibinfo {pages} {47} (\bibinfo {year}
		{1992})}\BibitemShut {NoStop}%
	\bibitem [{\citenamefont {Port{\'e}-Agel}\ \emph {et~al.}(2000)\citenamefont
		{Port{\'e}-Agel}, \citenamefont {Meneveau},\ and\ \citenamefont
		{Parlange}}]{porte2000scale}%
	\BibitemOpen
	\bibfield  {author} {\bibinfo {author} {\bibfnamefont {F.}~\bibnamefont
			{Port{\'e}-Agel}}, \bibinfo {author} {\bibfnamefont {C.}~\bibnamefont
			{Meneveau}},\ and\ \bibinfo {author} {\bibfnamefont {M.~B.}\ \bibnamefont
			{Parlange}},\ }\bibfield  {title} {\bibinfo {title} {A scale-dependent
			dynamic model for large-eddy simulation: application to a neutral atmospheric
			boundary layer},\ }\href@noop {} {\bibfield  {journal} {\bibinfo  {journal}
			{J. Fluid Mech.}\ }\textbf {\bibinfo {volume} {415}},\ \bibinfo {pages} {261}
		(\bibinfo {year} {2000})}\BibitemShut {NoStop}%
	\bibitem [{\citenamefont {Bou-Zeid}\ \emph {et~al.}(2004)\citenamefont
		{Bou-Zeid}, \citenamefont {Meneveau},\ and\ \citenamefont
		{Parlange}}]{bou2004large}%
	\BibitemOpen
	\bibfield  {author} {\bibinfo {author} {\bibfnamefont {E.}~\bibnamefont
			{Bou-Zeid}}, \bibinfo {author} {\bibfnamefont {C.}~\bibnamefont {Meneveau}},\
		and\ \bibinfo {author} {\bibfnamefont {M.~B.}\ \bibnamefont {Parlange}},\
	}\bibfield  {title} {\bibinfo {title} {Large-eddy simulation of neutral
			atmospheric boundary layer flow over heterogeneous surfaces: blending height
			and effective surface roughness},\ }\href@noop {} {\bibfield  {journal}
		{\bibinfo  {journal} {Water Resour. Res.}\ }\textbf {\bibinfo {volume} {40}}
		(\bibinfo {year} {2004})}\BibitemShut {NoStop}%
	\bibitem [{\citenamefont {Huang}\ \emph
		{et~al.}(2009{\natexlab{b}})\citenamefont {Huang}, \citenamefont {Cassiani},\
		and\ \citenamefont {Albertson}}]{huang2009effects}%
	\BibitemOpen
	\bibfield  {author} {\bibinfo {author} {\bibfnamefont {J.}~\bibnamefont
			{Huang}}, \bibinfo {author} {\bibfnamefont {M.}~\bibnamefont {Cassiani}},\
		and\ \bibinfo {author} {\bibfnamefont {J.}~\bibnamefont {Albertson}},\
	}\bibfield  {title} {\bibinfo {title} {The effects of vegetation density on
			coherent turbulent structures within the canopy sublayer: A large-eddy
			simulation study},\ }\href@noop {} {\bibfield  {journal} {\bibinfo  {journal}
			{Boundary-layer meteorol.}\ }\textbf {\bibinfo {volume} {133}},\ \bibinfo
		{pages} {253} (\bibinfo {year} {2009}{\natexlab{b}})}\BibitemShut {NoStop}%
	\bibitem [{\citenamefont {Chamecki}\ \emph {et~al.}(2009)\citenamefont
		{Chamecki}, \citenamefont {Meneveau},\ and\ \citenamefont
		{Parlange}}]{chamecki2009large}%
	\BibitemOpen
	\bibfield  {author} {\bibinfo {author} {\bibfnamefont {M.}~\bibnamefont
			{Chamecki}}, \bibinfo {author} {\bibfnamefont {C.}~\bibnamefont {Meneveau}},\
		and\ \bibinfo {author} {\bibfnamefont {M.~B.}\ \bibnamefont {Parlange}},\
	}\bibfield  {title} {\bibinfo {title} {Large eddy simulation of pollen
			transport in the atmospheric boundary layer},\ }\href@noop {} {\bibfield
		{journal} {\bibinfo  {journal} {J. Aerosol Sci.}\ }\textbf {\bibinfo {volume}
			{40}},\ \bibinfo {pages} {241} (\bibinfo {year} {2009})}\BibitemShut
	{NoStop}%
	\bibitem [{\citenamefont {Lu}\ and\ \citenamefont
		{Port{\'e}-Agel}(2010)}]{lu2010modulated}%
	\BibitemOpen
	\bibfield  {author} {\bibinfo {author} {\bibfnamefont {H.}~\bibnamefont
			{Lu}}\ and\ \bibinfo {author} {\bibfnamefont {F.}~\bibnamefont
			{Port{\'e}-Agel}},\ }\bibfield  {title} {\bibinfo {title} {A modulated
			gradient model for large-eddy simulation: application to a neutral
			atmospheric boundary layer},\ }\href@noop {} {\bibfield  {journal} {\bibinfo
			{journal} {Phys. Fluids}\ }\textbf {\bibinfo {volume} {22}},\ \bibinfo
		{pages} {015109} (\bibinfo {year} {2010})}\BibitemShut {NoStop}%
	\bibitem [{\citenamefont {Calaf}\ \emph {et~al.}(2010)\citenamefont {Calaf},
		\citenamefont {Meneveau},\ and\ \citenamefont {Meyers}}]{calaf2010large}%
	\BibitemOpen
	\bibfield  {author} {\bibinfo {author} {\bibfnamefont {M.}~\bibnamefont
			{Calaf}}, \bibinfo {author} {\bibfnamefont {C.}~\bibnamefont {Meneveau}},\
		and\ \bibinfo {author} {\bibfnamefont {J.}~\bibnamefont {Meyers}},\
	}\bibfield  {title} {\bibinfo {title} {Large eddy simulation study of fully
			developed wind-turbine array boundary layers},\ }\href@noop {} {\bibfield
		{journal} {\bibinfo  {journal} {Phys. Fluids}\ }\textbf {\bibinfo {volume}
			{22}},\ \bibinfo {pages} {015110} (\bibinfo {year} {2010})}\BibitemShut
	{NoStop}%
	\bibitem [{\citenamefont {Fang}\ and\ \citenamefont
		{Port{\'e}-Agel}(2015)}]{fang2015large}%
	\BibitemOpen
	\bibfield  {author} {\bibinfo {author} {\bibfnamefont {J.}~\bibnamefont
			{Fang}}\ and\ \bibinfo {author} {\bibfnamefont {F.}~\bibnamefont
			{Port{\'e}-Agel}},\ }\bibfield  {title} {\bibinfo {title} {Large-eddy
			simulation of very-large-scale motions in the neutrally stratified
			atmospheric boundary layer},\ }\href@noop {} {\bibfield  {journal} {\bibinfo
			{journal} {Boundary-Layer Meteorol.}\ }\textbf {\bibinfo {volume} {155}},\
		\bibinfo {pages} {397} (\bibinfo {year} {2015})}\BibitemShut {NoStop}%
	\bibitem [{\citenamefont {Sillero}\ \emph {et~al.}(2014)\citenamefont
		{Sillero}, \citenamefont {Jim{\'e}nez},\ and\ \citenamefont
		{Moser}}]{sillero2014two}%
	\BibitemOpen
	\bibfield  {author} {\bibinfo {author} {\bibfnamefont {J.~A.}\ \bibnamefont
			{Sillero}}, \bibinfo {author} {\bibfnamefont {J.}~\bibnamefont
			{Jim{\'e}nez}},\ and\ \bibinfo {author} {\bibfnamefont {R.~D.}\ \bibnamefont
			{Moser}},\ }\bibfield  {title} {\bibinfo {title} {Two-point statistics for
			turbulent boundary layers and channels at {Reynolds} numbers up to
			$\delta^+\approx 2000$},\ }\href@noop {} {\bibfield  {journal} {\bibinfo
			{journal} {Phys. Fluids}\ }\textbf {\bibinfo {volume} {26}},\ \bibinfo
		{pages} {105109} (\bibinfo {year} {2014})}\BibitemShut {NoStop}%
	\bibitem [{\citenamefont {Meyers}(2011)}]{meyers2011error}%
	\BibitemOpen
	\bibfield  {author} {\bibinfo {author} {\bibfnamefont {J.}~\bibnamefont
			{Meyers}},\ }\bibfield  {title} {\bibinfo {title} {Error-landscape assessment
			of large-eddy simulations: a review of the methodology},\ }\href@noop {}
	{\bibfield  {journal} {\bibinfo  {journal} {J. Sci. Comput.}\ }\textbf
		{\bibinfo {volume} {49}},\ \bibinfo {pages} {65} (\bibinfo {year}
		{2011})}\BibitemShut {NoStop}%
	\bibitem [{\citenamefont {Munters}\ \emph {et~al.}(2016)\citenamefont
		{Munters}, \citenamefont {Meneveau},\ and\ \citenamefont
		{Meyers}}]{munters2016shifted}%
	\BibitemOpen
	\bibfield  {author} {\bibinfo {author} {\bibfnamefont {W.}~\bibnamefont
			{Munters}}, \bibinfo {author} {\bibfnamefont {C.}~\bibnamefont {Meneveau}},\
		and\ \bibinfo {author} {\bibfnamefont {J.}~\bibnamefont {Meyers}},\
	}\bibfield  {title} {\bibinfo {title} {Shifted periodic boundary conditions
			for simulations of wall-bounded turbulent flows},\ }\href@noop {} {\bibfield
		{journal} {\bibinfo  {journal} {Phys. Fluids}\ }\textbf {\bibinfo {volume}
			{28}},\ \bibinfo {pages} {025112} (\bibinfo {year} {2016})}\BibitemShut
	{NoStop}%
	\bibitem [{\citenamefont {Bou-Zeid}\ \emph {et~al.}(2005)\citenamefont
		{Bou-Zeid}, \citenamefont {Meneveau},\ and\ \citenamefont
		{Parlange}}]{bou2005scale}%
	\BibitemOpen
	\bibfield  {author} {\bibinfo {author} {\bibfnamefont {E.}~\bibnamefont
			{Bou-Zeid}}, \bibinfo {author} {\bibfnamefont {C.}~\bibnamefont {Meneveau}},\
		and\ \bibinfo {author} {\bibfnamefont {M.}~\bibnamefont {Parlange}},\
	}\bibfield  {title} {\bibinfo {title} {A scale-dependent {Lagrangian} dynamic
			model for large eddy simulation of complex turbulent flows},\ }\href@noop {}
	{\bibfield  {journal} {\bibinfo  {journal} {Phys. Fluids}\ }\textbf {\bibinfo
			{volume} {17}},\ \bibinfo {pages} {025105} (\bibinfo {year}
		{2005})}\BibitemShut {NoStop}%
	\bibitem [{\citenamefont {Smagorinsky}(1963)}]{smagorinsky1963general}%
	\BibitemOpen
	\bibfield  {author} {\bibinfo {author} {\bibfnamefont {J.}~\bibnamefont
			{Smagorinsky}},\ }\bibfield  {title} {\bibinfo {title} {General circulation
			experiments with the primitive equations: I. the basic experiment},\
	}\href@noop {} {\bibfield  {journal} {\bibinfo  {journal} {Mon. Weather
				Rev.}\ }\textbf {\bibinfo {volume} {91}},\ \bibinfo {pages} {99} (\bibinfo
		{year} {1963})}\BibitemShut {NoStop}%
	\bibitem [{\citenamefont {Mason}\ and\ \citenamefont
		{Thomson}(1992)}]{Mason1992}%
	\BibitemOpen
	\bibfield  {author} {\bibinfo {author} {\bibfnamefont {P.~J.}\ \bibnamefont
			{Mason}}\ and\ \bibinfo {author} {\bibfnamefont {D.~J.}\ \bibnamefont
			{Thomson}},\ }\bibfield  {title} {\bibinfo {title} {Stochastic backscatter in
			large-eddy simulations of boundary layers},\ }\href@noop {} {\bibfield
		{journal} {\bibinfo  {journal} {J. Fluid Mech.}\ }\textbf {\bibinfo {volume}
			{242}},\ \bibinfo {pages} {51} (\bibinfo {year} {1992})}\BibitemShut
	{NoStop}%
	\bibitem [{\citenamefont {Stevens}\ \emph {et~al.}(2014)\citenamefont
		{Stevens}, \citenamefont {Wilczek},\ and\ \citenamefont
		{Meneveau}}]{stevens2014large}%
	\BibitemOpen
	\bibfield  {author} {\bibinfo {author} {\bibfnamefont {R.~J.}\ \bibnamefont
			{Stevens}}, \bibinfo {author} {\bibfnamefont {M.}~\bibnamefont {Wilczek}},\
		and\ \bibinfo {author} {\bibfnamefont {C.}~\bibnamefont {Meneveau}},\
	}\bibfield  {title} {\bibinfo {title} {Large-eddy simulation study of the
			logarithmic law for second-and higher-order moments in turbulent wall-bounded
			flow},\ }\href@noop {} {\bibfield  {journal} {\bibinfo  {journal} {J. Fluid
				Mech.}\ }\textbf {\bibinfo {volume} {757}},\ \bibinfo {pages} {888} (\bibinfo
		{year} {2014})}\BibitemShut {NoStop}%
	\bibitem [{\citenamefont {Canuto}\ \emph {et~al.}(1988)\citenamefont {Canuto},
		\citenamefont {Quarteroni}, \citenamefont {Hussaini},\ and\ \citenamefont
		{Zang}}]{canuto1988spectral}%
	\BibitemOpen
	\bibfield  {author} {\bibinfo {author} {\bibfnamefont {C.}~\bibnamefont
			{Canuto}}, \bibinfo {author} {\bibfnamefont {A.}~\bibnamefont {Quarteroni}},
		\bibinfo {author} {\bibfnamefont {M.~Y.}\ \bibnamefont {Hussaini}},\ and\
		\bibinfo {author} {\bibfnamefont {T.~A.}\ \bibnamefont {Zang}},\ }\href@noop
	{} {\emph {\bibinfo {title} {Spectral methods in fluid dynamics}}}\ (\bibinfo
	{publisher} {Berlin},\ \bibinfo {year} {1988})\BibitemShut {NoStop}%
	\bibitem [{\citenamefont {Verstappen}\ and\ \citenamefont
		{Veldman}(2003)}]{verstappen2003symmetry}%
	\BibitemOpen
	\bibfield  {author} {\bibinfo {author} {\bibfnamefont {R.}~\bibnamefont
			{Verstappen}}\ and\ \bibinfo {author} {\bibfnamefont {A.}~\bibnamefont
			{Veldman}},\ }\bibfield  {title} {\bibinfo {title} {Symmetry-preserving
			discretization of turbulent flow},\ }\href@noop {} {\bibfield  {journal}
		{\bibinfo  {journal} {J. Comput. Phys.}\ }\textbf {\bibinfo {volume} {187}},\
		\bibinfo {pages} {343} (\bibinfo {year} {2003})}\BibitemShut {NoStop}%
	\bibitem [{\citenamefont
		{Sirovich}(1987{\natexlab{b}})}]{sirovich1987turbulenceII}%
	\BibitemOpen
	\bibfield  {author} {\bibinfo {author} {\bibfnamefont {L.}~\bibnamefont
			{Sirovich}},\ }\bibfield  {title} {\bibinfo {title} {Turbulence and the
			dynamics of coherent structures. {II}. symmetries and transformations},\
	}\href@noop {} {\bibfield  {journal} {\bibinfo  {journal} {Q. Appl. Math}\
		}\textbf {\bibinfo {volume} {45}},\ \bibinfo {pages} {573} (\bibinfo {year}
		{1987}{\natexlab{b}})}\BibitemShut {NoStop}%
	\bibitem [{\citenamefont {Perry}\ \emph {et~al.}(1986)\citenamefont {Perry},
		\citenamefont {Henbest},\ and\ \citenamefont {Chong}}]{perry1986theoretical}%
	\BibitemOpen
	\bibfield  {author} {\bibinfo {author} {\bibfnamefont {A.}~\bibnamefont
			{Perry}}, \bibinfo {author} {\bibfnamefont {S.}~\bibnamefont {Henbest}},\
		and\ \bibinfo {author} {\bibfnamefont {M.}~\bibnamefont {Chong}},\ }\bibfield
	{title} {\bibinfo {title} {A theoretical and experimental study of wall
			turbulence},\ }\href@noop {} {\bibfield  {journal} {\bibinfo  {journal} {J.
				Fluid Mech.}\ }\textbf {\bibinfo {volume} {165}},\ \bibinfo {pages} {163}
		(\bibinfo {year} {1986})}\BibitemShut {NoStop}%
	\bibitem [{\citenamefont {Knight}\ and\ \citenamefont
		{Sirovich}(1990)}]{knight1990kolmogorov}%
	\BibitemOpen
	\bibfield  {author} {\bibinfo {author} {\bibfnamefont {B.}~\bibnamefont
			{Knight}}\ and\ \bibinfo {author} {\bibfnamefont {L.}~\bibnamefont
			{Sirovich}},\ }\bibfield  {title} {\bibinfo {title} {Kolmogorov inertial
			range for inhomogeneous turbulent flows},\ }\href@noop {} {\bibfield
		{journal} {\bibinfo  {journal} {Phys. Rev. Lett.}\ }\textbf {\bibinfo
			{volume} {65}},\ \bibinfo {pages} {1356} (\bibinfo {year}
		{1990})}\BibitemShut {NoStop}%
	\bibitem [{\citenamefont {Moser}(1994)}]{moser1994kolmogorov}%
	\BibitemOpen
	\bibfield  {author} {\bibinfo {author} {\bibfnamefont {R.~D.}\ \bibnamefont
			{Moser}},\ }\bibfield  {title} {\bibinfo {title} {Kolmogorov inertial range
			spectra for inhomogeneous turbulence},\ }\href@noop {} {\bibfield  {journal}
		{\bibinfo  {journal} {Phys. Fluids}\ }\textbf {\bibinfo {volume} {6}},\
		\bibinfo {pages} {794} (\bibinfo {year} {1994})}\BibitemShut {NoStop}%
	\bibitem [{\citenamefont {Chapman}(1979)}]{chapman1979computational}%
	\BibitemOpen
	\bibfield  {author} {\bibinfo {author} {\bibfnamefont {D.~R.}\ \bibnamefont
			{Chapman}},\ }\bibfield  {title} {\bibinfo {title} {Computational
			aerodynamics development and outlook},\ }\href@noop {} {\bibfield  {journal}
		{\bibinfo  {journal} {AIAA J.}\ }\textbf {\bibinfo {volume} {17}},\ \bibinfo
		{pages} {1293} (\bibinfo {year} {1979})}\BibitemShut {NoStop}%
	\bibitem [{\citenamefont {Hutchins}\ \emph {et~al.}(2012)\citenamefont
		{Hutchins}, \citenamefont {Chauhan}, \citenamefont {Marusic}, \citenamefont
		{Monty},\ and\ \citenamefont {Klewicki}}]{hutchins2012towards}%
	\BibitemOpen
	\bibfield  {author} {\bibinfo {author} {\bibfnamefont {N.}~\bibnamefont
			{Hutchins}}, \bibinfo {author} {\bibfnamefont {K.}~\bibnamefont {Chauhan}},
		\bibinfo {author} {\bibfnamefont {I.}~\bibnamefont {Marusic}}, \bibinfo
		{author} {\bibfnamefont {J.}~\bibnamefont {Monty}},\ and\ \bibinfo {author}
		{\bibfnamefont {J.}~\bibnamefont {Klewicki}},\ }\bibfield  {title} {\bibinfo
		{title} {Towards reconciling the large-scale structure of turbulent boundary
			layers in the atmosphere and laboratory},\ }\href@noop {} {\bibfield
		{journal} {\bibinfo  {journal} {Boundary-Layer Meteorol.}\ }\textbf {\bibinfo
			{volume} {145}},\ \bibinfo {pages} {273} (\bibinfo {year}
		{2012})}\BibitemShut {NoStop}%
	\bibitem [{\citenamefont {Marusic}\ \emph {et~al.}(2013)\citenamefont
		{Marusic}, \citenamefont {Monty}, \citenamefont {Hultmark},\ and\
		\citenamefont {Smits}}]{marusic2013logarithmic}%
	\BibitemOpen
	\bibfield  {author} {\bibinfo {author} {\bibfnamefont {I.}~\bibnamefont
			{Marusic}}, \bibinfo {author} {\bibfnamefont {J.~P.}\ \bibnamefont {Monty}},
		\bibinfo {author} {\bibfnamefont {M.}~\bibnamefont {Hultmark}},\ and\
		\bibinfo {author} {\bibfnamefont {A.~J.}\ \bibnamefont {Smits}},\ }\bibfield
	{title} {\bibinfo {title} {On the logarithmic region in wall turbulence},\
	}\href@noop {} {\bibfield  {journal} {\bibinfo  {journal} {J. Fluid Mech.}\
		}\textbf {\bibinfo {volume} {716}} (\bibinfo {year} {2013})}\BibitemShut
	{NoStop}%
	\bibitem [{\citenamefont {Meneveau}\ and\ \citenamefont
		{Marusic}(2013)}]{meneveau2013generalized}%
	\BibitemOpen
	\bibfield  {author} {\bibinfo {author} {\bibfnamefont {C.}~\bibnamefont
			{Meneveau}}\ and\ \bibinfo {author} {\bibfnamefont {I.}~\bibnamefont
			{Marusic}},\ }\bibfield  {title} {\bibinfo {title} {Generalized logarithmic
			law for high-order moments in turbulent boundary layers},\ }\href@noop {}
	{\bibfield  {journal} {\bibinfo  {journal} {J. Fluid Mech.}\ }\textbf
		{\bibinfo {volume} {719}} (\bibinfo {year} {2013})}\BibitemShut {NoStop}%
	\bibitem [{\citenamefont {Jim{\'e}nez}(2004)}]{jimenez2004turbulent}%
	\BibitemOpen
	\bibfield  {author} {\bibinfo {author} {\bibfnamefont {J.}~\bibnamefont
			{Jim{\'e}nez}},\ }\bibfield  {title} {\bibinfo {title} {Turbulent flows over
			rough walls},\ }\href@noop {} {\bibfield  {journal} {\bibinfo  {journal}
			{Annu. Rev. Fluid Mech.}\ }\textbf {\bibinfo {volume} {36}},\ \bibinfo
		{pages} {173} (\bibinfo {year} {2004})}\BibitemShut {NoStop}%
	\bibitem [{\citenamefont {Townsend}(1976)}]{townsend1976structure}%
	\BibitemOpen
	\bibfield  {author} {\bibinfo {author} {\bibfnamefont {A.}~\bibnamefont
			{Townsend}},\ }\bibfield  {title} {\bibinfo {title} {The structure of
			turbulent shear flow},\ }\href@noop {} {\bibfield  {journal} {\bibinfo
			{journal} {Cambridge and New York, Cambridge University Press, 1976. 438 p.}\
		} (\bibinfo {year} {1976})}\BibitemShut {NoStop}%
\end{thebibliography}

\providecommand{\noopsort}[1]{}\providecommand{\singleletter}[1]{#1}%

\end{document}